\documentstyle[psfig,times]{mn}
 
\def\etal{{\it et al.~}}

\def\la{\hbox{ \raise.35ex\rlap{$<$}\lower.6ex\hbox{$\sim$}\ }}
\def\ga{\hbox{ \raise.35ex\rlap{$>$}\lower.6ex\hbox{$\sim$}\ }}

\def\W2{{\cal W}}

\begin{document}
 
\title[Galactosynthesis: Halo Histories, Star Formation, and
Disks]{Galactosynthesis: Halo Histories, Star Formation, and Disks}  
 
\author[Buchalter, Jimenez $\&$ Kamionkowski]
{Ari Buchalter$^1$, Raul Jimenez$^2$ $\&$ Marc Kamionkowski$^1$ \\
$^1$Theoretical Astrophysics Group, Mail Code 130-33, 
California Institute of Technology, Pasadena, CA
91125. \\
ari@tapir.caltech.edu, kamion@tapir.caltech.edu\\
$^2$Institute for Astronomy, University of Edinburgh, Blackford Hill, 
Edinburgh EH9 3HJ, UK. \\
raul@roe.ac.uk}

\maketitle

\begin{abstract}         

We investigate the effects of a variety of ingredients that must enter into a
realistic model for disk-galaxy formation, focusing primarily on the
Tully-Fisher (TF) relation and its scatter in several wavebands.  In
particular, we employ analytic distributions for halo-formation redshifts and
halo spins, empirical star-formation rates and initial mass functions,
realistic stellar populations, and chemical evolution of the gas.  Our main
findings are: (a) the slope, normalization, and scatter of the TF relation
across various wavebands is determined {\em both} by halo properties and star
formation in the disk; (b) TF scatter owes primarily to the spread in
formation redshifts.  The scatter can be measurably reduced by chemical
evolution, and also in some cases by the weak anti-correlation between peak
height and spin; (c) multi-wavelength constraints can be important in
distinguishing between models which appear to fit the TF relation in $I$ or
$K$; (d) successful models seem to require that the bulk of disk formation
cannot occur too early ($z>2$) or too late ($z<0.5$), and are inconsistent
with high values of $\Omega_0$; (e) a realistic model with the above
ingredients can reasonably reproduce the observed $z=0$ TF relation in {\em
all} bands ($B$, $R$, $I$, and $K$). It can also account for the $z=1$
$B$-band TF relation and yield rough agreement with the local $B$ and $K$
luminosity functions and $B$-band surface-brightness--magnitude relation.  In
such a model, the near-infrared TF relation at $z=1$ is similar to that at
$z=0$, while bluer bands show a markedly steeper TF slope at high redshift.
The remarkable agreement with observations suggests that the amount of gas
that is expelled or poured into a disk galaxy must be small (though small
fluctuations might serve to better align $B$-band predictions with
observations), and that the specific angular momentum of the baryons must
roughly equal that of the halo; there is little room for angular momentum
transfer. In an appendix we present analytic fits to stellar-population
synthesis models.

\end{abstract}

 
\section{Introduction} 

Spiral galaxies are particularly important in the study of galaxy formation,
as they are believed to undergo a relatively smooth formation process, and
serve as the building blocks in the formation of other galactic systems
through mergers. Thus, spiral galaxies should be the easiest to model and
should provide clues as to the basic physics underlying galaxy formation.
Various lines of observational evidence serve to guide our understanding of
spirals, including their measured luminosity function (LF), surface-brightness
distribution, star-formation history, chemical composition, and dynamical
properties. Of particular significance is the Tully-Fisher (TF) relation, a
remarkably tight correlation between the luminosity and rotation speed of
spirals. For a given wavelength, $\lambda$, the TF relation obeys the form
$L_\lambda=A_\lambda {V_c}^{\gamma_\lambda}$, or
\begin{equation}
M_\lambda=a_\lambda + b_\lambda \log{V_c},
\label{tf} 
\end{equation}
where $M_\lambda$ is the absolute magnitude\footnote{A subscripted
$M$ is understood to refer to absolute magnitude in a given band,
while an $M$ without a subscript refers to mass.},
and $b_\lambda=-2.5\gamma_\lambda$ is the slope of the
relation.
The debate continues as to whether this
relationship results primarily from initial conditions, i.e., from the
properties of the parent halo (Dalcanton, Spergel, \& Summers 1997;
Mo, Mao, \& White 1998; Firmani \& Avila-Reese 1998;
Avila-Reese, Firmani, \& Hernandez 1998; Navarro \& Steinmetz 2000; 
Mo \& Mao 2000), 
self-regulating feedback processes associated with star formation in the disk 
(Silk 1997), or a combination of both (Heavens \& 
Jimenez 1999; Somerville \& Primack 1999; van den Bosch 2000).

A TF relation arises quite naturally if one simply assumes that
galactic halos formed at roughly the same time and that 
luminosity is proportional to the baryonic mass, which is in turn 
proportional to the halo mass.  More realistically, the
luminosity may depend on the galactic spin, as disks formed in
high-spin halos will be larger and thinner, and thus lead to
lower star-formation rates.  Still, even with spin, a TF
relation arises if halos all formed at roughly the same epoch.
In both cases (with and without spin), scatter in the redshift
of halo formation should lead to scatter in the central
densities of the halos and thus to scatter in the TF relation.

Several groups have made considerable progress in understanding spiral-galaxy
properties along these lines using semi-analytical models (SAMs).  Eisenstein
\& Loeb (1996, hereafter EL96) used Monte-Carlo realizations of halo-formation
histories to calculate the minimum TF scatter which should arise from the
spread in halo-formation times. They concluded that unless spirals form at $z
\ga 1$, without subsequently accreting much mass, the TF relation cannot arise
simply from initial conditions, but must instead owe to some feedback
mechanism which decouples the luminosity from the halo history.

More recently, other groups have investigated detailed SAMs of disk-galaxy
formation which incorporate such features as formation histories derived
directly from $N$-body simulations, universal halo profiles, adiabatic disk
contraction, bulge formation via stability criteria, star formation, supernova
feedback, dust, cooling, and mergers (Somerville \& Primack 1999; van den
Bosch 2000; Firmani \& Avila-Reese 1998; Navarro \& Steinmetz 2000).  Their
conclusions differ as to the relative importance of initial conditions vs.
feedback from star formation and/or supernovae in defining the TF relation.
These studies generally agree, however, that the spread in halo-formation
redshifts is a significant source of scatter in the TF relation and that
reconciling models with TF observations seems to require a low matter density
($\Omega_0 \sim 0.3$ -- 0.5) and disk formation at high redshift ($z \ga 1$).

One powerful test of such models, which has not been fully appreciated, is the
simultaneous comparison of their predictions to TF data from {\em several}
wavebands.  Some previous work has considered only an assumed value for the
mass-to-light ratio in a given band, rather than using stellar population
models to predict broad-band magnitudes, and many authors have investigated TF
predictions for only a single waveband, typically $I$ or $K$, where the
observed TF scatter is the smallest [$\sim 0.4$ mag in the most carefully
defined samples (Willick \etal 1995, 1996, 1997; Tully \etal 1998)].  Since
these wavelengths measure primarily the oldest, shell-burning, stellar
populations, such measurements are, by construction, effectively probing only
the total mass of the galaxy. Thus, while some authors have claimed success in
fitting the near-IR TF relation, their predictions in bluer bands, where many
model ingredients would be most strongly manifested owing to the younger
populations probed, have gone unchecked.  With realistic stellar-population
models, the leverage gained by spanning a range of wavelengths should
therefore be crucial in distinguishing models which produce similar
near-infrared TF predictions, and help to assess the importance of various SAM
features.  Heavens and Jimenez (1999; hereafter HJ99), used a simple halo and
disk model combined with empirical star-formation properties to investigate
the role of star formation in the TF relation, did examine the simultaneous
constraints from various wavebands, but did not address some of the other key
considerations listed above.

Our approach here will focus on constructing, as much as possible, a model
``from the ground up,'' i.e., starting with a minimal set of simple,
well-motivated assumptions, and individually investigating the impact of
various plausible modifications.  In essence, we are asking whether a
plausible model with fewer free parameters\footnote{Most of the freedom in our
models comes from the cosmological-parameter choices, which we generally fix
to be consistent with current estimates; other choices for these parameters
would yield other TF relations, usually inconsistent with the data.}  than are
usually incorporated into current SAMs can reasonably pass various
observational tests.  Though the resulting model will lack many of the
sophisticated features of other SAMs and will surely be an oversimplification
in many respects, it may shed important light on the issue of which physical
ingredients are truly essential in determining certain disk properties.

We build on the work of HJ99, using their halo/disk model and stellar
populations code, but generalizing it to include a variety of important
features, such as analytic models for the distribution of halo-formation
redshifts and spins (as well as the predicted anti-correlation between the
two). This allows us to assess not only the impact of these distributions on
TF predictions, but the dependence on cosmological parameters and the power
spectrum as well. We also include for the first time an analytic description
of chemical evolution. This measurably reduces the scatter in the TF relation,
which is found to arise mainly from the spread in formation times.  The
overall result will be a model for disk galaxies which passes a host of
observational tests, reproducing the observed TF relation, locally and at
$z=1$, in all relevant wavebands {\em with the right magnitude of scatter.}
It also succeeds in roughly fitting the observed surface-brightness
distribution of spirals and, within its limitations, reasonably reproduces the
observed $B$- and $K$-band luminosity functions.  The model necessitates
cosmological parameter values which are in line with current estimates, and
requires that the bulk of disk formation occur in the range $0.5 < z < 2$.
This agreement implies that, although plausible, other mechanisms which would
add or remove gas from the disk (such as mergers, gas expulsion by supernovae,
etc.), and thereby distort the predicted luminosity, are not necessarily
required to successfully predict these global disk-galaxy properties.

\section{The Model}

\subsection{Our Starting Point}

We build on the skeletal disk-galaxy model of HJ99. We review the basic
ingredients here and refer the reader to HJ99 for a more complete description.
This model assumes that after dark halos separate from the Hubble flow and
collapse, they subsequently relax to an isothermal sphere of mass $M$, and the
baryons instantaneously settle into a central disk with mass $M_d=m_d M$,
where $m_d=\Omega_b/\Omega_0$, $\Omega_0$ is the present-day non-relativistic
matter density in units of the critical density, and the baryon density,
$\Omega_b$, is set by nucleosynthesis, $\Omega_b=0.019 h^{-2}$ (Tytler \etal
1999).  The disk is assumed to have an exponential profile,
$\Sigma(R)=\Sigma_0 \exp(-R/R_d)$, with central surface density $\Sigma_0$ and
scale length $R_d$, such that $M_d=2 \pi \Sigma_0 {R_d}^2$.  Since the
assumptions of a singular isothermal sphere and instantaneous disk formation
cannot strictly be true, the apparent success of the model will argue against
any strong influence of details such as the profile shape (Mo \& Mao 2000) or
disk-contraction process on the resultant properties of spirals.

With the gaseous disk in place at the time of formation, star formation
proceeds according to an empirical Schmidt law (Kennicutt 1998), such that the
star-formation rate (SFR) depends only on the local gas surface density. Given
the conservation of baryonic mass, the total SFR, as well as remaining gas
fraction, can then be obtained as functions of time since formation by
integrating over the disk [see equations (3)--(9) in HJ99].  The model ignores
any gas returned to the ISM by stars as well as late infall of fresh gas.

Using the SFR and an assumed stellar initial mass function (IMF), the
properties of the stellar populations formed at each time interval are
calculated using the spectrophotometric stellar-evolution code of Jimenez
\etal (1998, 2000), and broad-band magnitudes derived from the resulting
spectra are integrated over the life of the galaxy to yield luminosities in
various bands.  Other stellar-population codes generally yield results which
agree to within 0.1--0.2 mag and have comparable dispersions (Jimenez \etal
2000). Since we are integrating the SFR over time, weighted by the specific
luminosities of model stellar populations as a function of age, metallicity,
spin, etc., the model can predict disk luminosities and surface brightnesses
(in a variety of wavebands) at any point in their evolution.  More details
about the calculation of the luminosities of the stellar populations, as well
as some possibly useful analytic fitting formulas, are provided in an
Appendix.

To summarize, for a given set of cosmological parameters, the model of HJ99
takes as input a total galactic mass, $M$, and spin, $\lambda$, and can
output, as functions of time, the disk luminosity and surface brightness in
various wavebands, as well as the metallicity.

\subsection{Spherical-Collapse Model}

We will be interested in the relation between the size and mass of a galactic
halo, and this is fixed by the formation redshift---halos of a given mass that
undergo collapse earlier are expected to be smaller and denser---through the
spherical-collapse model.  HJ99 used a simple approximation for the relation
between the mass, size, and collapse redshift, neglecting the full impact of a
possible cosmological constant.  Since our aim is to carefully study the
effects of variations in the formation-redshift distribution, we treat the
spherical-collapse model more precisely, following the approach of Wang \&
Steinhardt (1998).

The model will of course depend on the background cosmogony, which effectively
sets the initial conditions and dictates the course of structure formation. We
consider various cold-dark-matter (CDM) models with a power spectrum given by
$P(k,z) = A D^2(z) k^n T^2(k)$, where $A$ is the overall amplitude, $D(z)$ is
the linear growth factor, and $k^n$ is the primordial power spectrum (we shall
assume an untilted, $n=1$ primordial spectrum throughout).  For the transfer
function, $T(k)$, we adopt the functional form of Bardeen \etal (1986),
parameterized by the shape parameter $\Gamma$.

We restrict out study to flat geometries ($\Omega_0 + \Omega_{\Lambda} = 1$,
where $\Omega_{\Lambda}$ is the energy density associated with the
cosmological constant, in units of the critical density), and consider various
values of $h$, the Hubble parameter ($H_0$) in units of 100 km s$^{-1}$
Mpc$^{-1}$.  The presence of a cosmological constant introduces several
modifications to the standard spherical-collapse model derived for the
Einstein-de Sitter case, which are often overlooked.  These include a nonzero
surface-pressure term in the virial relation, which in turn alters the
mass-circular velocity relationship, and introduces $\Omega_0$-dependent
corrections to $\rho_{coll}$--- the density of virialized objects, in units of
the present background density, given by a value of $\sim 179$ at $z=0$ in the
Einstein-de Sitter (EdS) case--- and to $\delta_c$, the overdensity ($\delta =
\rho/\overline{\rho}-1$) threshold for collapse, typically taken to be the
linearly-extrapolated value of $\sim 1.69$ in the EdS case (see equations (2),
(5), (6), and (7) and related discussion in Wang \& Steinhardt 1998).  We do
not reproduce all of these results, but merely write the equation for the
quantity which interests us directly, namely the circular velocity of the
disk, which can be related to the virial velocity of particles in the halo:
\begin{eqnarray}
{V_c}^2 & = &\frac{2}{3}\overline{{V_{\rm vir}}^2}  \\ \nonumber
\overline{{V_{\rm vir}}^2} & = &\frac{3}{5}\left[GMH_0\sqrt{\Omega_0\Delta_c/2}\right]^{2/3}
(1+z) \times \\ \nonumber
 & &  \left[1-\frac{2}{\Delta_c}\frac{1-\Omega_0}{\Omega_0 (1+z)^3}\right],
\label{V_c}
\end{eqnarray}
where $\Delta_c=\rho_{coll}/\rho_b$ is the $\Omega_0$-dependent ratio of the
halo to background density at collapse. We stress that these results can, in
general, differ substantially from their EdS counterparts.

We incorporate these $\Lambda$CDM generalizations into the Press-Schechter
(PS) theory, which will be our starting point for calculating halo abundances
and formation times. The theory states the familiar result that the comoving
number density of halos in a mass interval $dM$ about $M$ at redshift $z$ is
given by
\begin {eqnarray}
n(M,z)dM & = & -\sqrt{\frac{2}{\pi}}\frac{\rho_b}{M}\frac{\delta_c}{\sigma^2(R,z)}\frac
{d\sigma(R,z)}{dM}\times \\ \nonumber 
& &\exp\left[-\frac{{\delta_c}^2}{2\sigma^2(R,z)}\right]dM,
\label{ps}
\end{eqnarray}
where $R^3=3M/4\pi\rho_b$, $\rho_b$ is the constant, comoving background
matter density, $\sigma(R,z)$ is the rms fractional density perturbation in
spheres of radius $R$, and $\delta_c$ is here understood to depend on redshift
via an empirically derived relation (L. Wang, private communication).

\subsection{Formation-Redshift Distribution}

In standard hierarchical models of structure formation, halos of a fixed mass
form over a range of redshifts.  HJ99 examined only fixed values for the
redshift of formation, $z_f$, for halos, but considered several values to
illustrate the effect of a spread in $z_f$.  EL96 performed Monte-Carlo
realizations of halo formation histories, using the merger-tree approach
(Lacey \& Cole 1993, 1994) and a spherical-accretion model, and concluded that
the minimum TF scatter resulting solely from the calculated spread in halo
formation redshifts is already uncomfortably larger than that observed. They
concluded that satisfying the upper limit of a $\sim 10$\% relative error in
velocity dispersion requires that $\Delta z_f/(1+z_f) \la 20$\%. Other workers
(e.g., Bullock \etal 1999) have looked at the scatter in formation redshifts
(or the closely-related scatter in ``concentration parameters'') obtained from
numerical simulations and found better agreement with the observed scatter.

Although numerical simulations should in principle provide the best way to
evaluate the formation-redshift distribution, these have limitations in
practice.  In particular, it is difficult to collect enough statistics to
determine the distributions for different masses and/or to determine how these
distributions depend on the cosmological parameters or the power spectrum.  As
the analytic distributions discussed below demonstrate, it {\it is} indeed to
be expected that the formation-redshift distribution (and thus resulting
scatter in the TF relation) should depend on the mass, cosmological
parameters, and power spectrum.  Thus, although they can provide some
order-of-magnitude estimates, results from numerical simulations can be
mis-applied in SAMs if they are determined, say, for one mass from a
simulation with a particular choice of cosmogony, and then applied to other
masses and/or cosmogonies. Moreover, analytic methods allow for more obvious
and direct insight into the dependence of the model on various parameter
choices and assumptions.

Therefore, we follow Viana \& Liddle (1996), and explore two plausible
analytic models for the distribution of halo formation redshifts. The first,
denoted as the `S' distribution, is that of Sasaki (1994), which simply uses
the PS formalism to calculate the formation rate of bound objects, weighted by
the probability of arriving at some later time without merging, under the
assumption that the destruction efficiency is independent of mass.  In this
model, the distribution of formation redshifts of halos with mass $M$ is given
by
\begin{equation}
\frac{dn_S}{dz_f}= -\frac{{\delta_c}^2}{\sigma^2(R,z_f)}\frac
{n(M,z_f)}{\sigma(R,0)}\frac{d\sigma(R,z_f)}{dz_f}.
\end{equation}
The S distribution has the advantage of being independent of how one defines a
new halo, but the assumption of self-similar merging might be questioned (see,
e.g., Percival, Miller \& Peacock 2000).

The second distribution, denoted as the `LC' distribution, addressed this
shortcoming by employing the merger-tree approach of Lacey \& Cole (1993,
1994). In this formalism, halos continually grow with time, and so one must
explicitly define the time at which a particular halo has come into
existence\footnote{This ambiguity arises because in the excursion-set
formalism, one follows a random walk of trajectories of the spatially-filtered
$\delta$ to the first up-crossing above some threshold value. The ``tagged''
mass is statistically defined as the expectation value for the mass of a halo
in which a given particle will end up, but this will in general differ from
the actual mass in a given realization. This ambiguity can lead to predictions
of negative probability densities at $z\sim0$ in models with primordial
power-spectrum indices $n>0$.}  (e.g., the epoch of mass doubling).  Lacey \&
Cole (1993) derived expressions for the expected $z_f$ distribution of halos
both using analytic counting arguments and a Monte-Carlo approach to generate
merger histories for $M(t)$ explicitly. They found that the former approach
yielded excellent agreement with $N$-body results, while the latter
overestimated halo ages (Lacey \& Cole 1994).  Their results did not depend
strongly on the value of the local slope of the power spectrum.  The $z_f$
distribution in this model is given by
\begin{equation}
\frac{dn_{LC}}{dz_f}= p(w(z_f))\frac{dw(z_f)}{dz_f},
\end{equation}
where
\begin{eqnarray}
p(w(z_f)) & = & 2w(z_f)(f^{-1}-1) {\mbox {erfc}}
\left(\frac{w(z_f)}{\sqrt{2}}\right) - \nonumber \\
& & \sqrt{\frac{2}{\pi}}(f^{-1}-2)\exp\left(-\frac{w^2(z_f)}{2}\right), \\
w(z_f) & = & \frac{\delta_c (\sigma(M,0)/\sigma(M,z_f)-1)}{\sqrt{\sigma^2(fM,0)-\sigma^2(M,0)}}, 
\end{eqnarray}
and $f$ is the fraction of halo mass assembled by formation redshift $z_f$.
Thus at fixed redshift, models with higher values of $f$ produce younger
galaxies. Lacey \& Cole adopted $f=0.5$, identifying the mass doubling epoch
as the nominal point at which halos are born, while Viana \& Liddle (1996),
looking specifically at the mass-temperature relation for galaxy clusters, use
a value of $f=0.75$.  Note that in any of these models, the dependence on the
power spectrum is only through $\sigma(M)$, the rms fractional density
perturbation for the mass scale $M$; the dependence on $\Omega_0$ and $h$
enter through the linear-theory growth factor and in the spherical-collapse
physics, and through the star-formation model by fixing ages, baryon fraction,
etc.

\begin{figure}
\centerline{
\psfig{figure=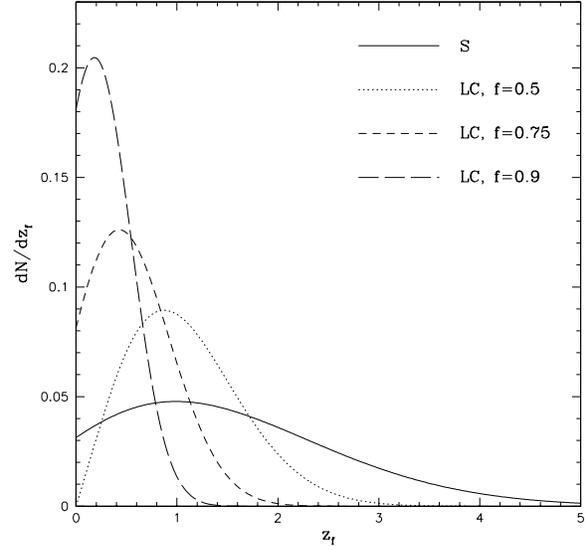,height=8cm,angle=0}}
\caption{\label{fig1} Formation-redshift distribution for the S distribution
(solid line) and for the LC distribution with $f=0.5$, 0.75, and 0.9 (dotted,
short-dashed, and long-dashed lines, respectively), assuming a $\Lambda$CDM
model with $\Omega_0=0.3$ and $h=0.65$.  Note the appreciable width of the S
distribution, and the decreasing width and peak $z_f$ values for LC
distributions with increasing $f$.}
\end{figure}

Figure \ref{fig1} depicts the distribution of formation redshifts for halos of
$10^{12} M_{\odot}$ both for the S distribution, and for the LC distribution
with $f$ values of 0.5, 0.75, and 0.9 for comparison. These results assume a
COBE-normalized $\Lambda$CDM model with $\Omega_0=0.3$ and $h=0.65$ (yielding,
in this case, a mean peak height of $\nu=0.69$, and $\sigma_8=1.08$).  Note
the large width of the S distribution, peaking at $z \sim 1.0$, but with a
significant tail extending out to $z \sim$ 4--5, and appreciable ongoing
formation today\footnote{This long tail will disappear if one uses the
formalism developed in Percival, Miller \& Peacock 2000}.  By contrast, LC
distributions with larger values of $f$ produce much narrower distributions
which peak at lower redshifts and fall more rapidly to high $z$.

\subsection{Spin Distributions}

Assuming constant specific angular momentum, the disk scale length can be
related to $\lambda$, the spin parameter of the halo (e.g., Mo, Mao, \& White
1998),
\begin{equation}
R_d = \frac{\lambda G M}{\sqrt{2} {V_c}^2}, 
\label{Rd}
\end{equation}
fixing the initial surface density of the gas and thus affecting the SFR,
luminosity, and surface brightness.  Therefore, it might be supposed that the
predicted TF relation and other properties would depend sensitively on spin.
HJ99 considered only fixed values for $\lambda$, rather than accounting for
its detailed distribution.  The spin distribution is usually taken to be a
log-normal centered at $\lambda\sim0.5$, as indicated by numerical
simulations.  However, as is the case for the formation-redshift distribution,
heuristic arguments suggest that the distribution of spin parameters should
depend on the mass, the formation redshift, and possibly on the cosmological
parameters and/or power spectrum.  If so, the effect of scatter in the spin
parameter on the TF relation could be different than that inferred from
numerical simulations.

Heavens \& Peacock (1988) calculated the distribution of tidal torques acting
on matter in the vicinity of mildly nonlinear density maxima, assuming a
spherical-accretion model to calculate binding energies, and derived the
resulting spin-parameter distribution.  They found an anti-correlation between
peak height and spin parameter, but pointed out that the intrinsically broad
range of the $\lambda$ distribution swamped the systematic shift with peak
height, $\nu$, resulting in a fairly weak anti-correlation. Catelan \& Theuns
(1996) extended these results to obtain the joint PDF for spin parameters and
peak masses using the distribution of peak shapes in different CDM
models. They confirmed the broadness of the spin-parameter distribution, and
the anti-correlation with peak height (which owes essentially to the fact that
higher peaks will generally be more spherical, and thus harder to spin
up). Their results, however, exhibit a systematic shift toward higher
$\lambda$ values than those of Heavens \& Peacock, such that $\nu=1$ peaks
have rms $\lambda$ values of $\sim 0.15$, instead of $0.05$, thus closer to
the observed value of 0.5 for spirals.  The anticorrelation conjecture has not
been fully tested by simulations, but some support comes from Ueda \etal
(1994).  Lemson \& Kauffmann (1999) argued against an environment-dependent
$\lambda$.  However, the environment was defined on a 10 $h^{-1}$ Mpc scale,
and we do not expect strong correlation between the density field smoothed on
galaxy scales and this large scale.

Since the indications for a distribution peaked at higher spins and an
anti-correlation with peak height are good, but not yet well-established, we
consider two models, with and without the joint distribution in $\lambda$ and
$\nu$. For the former, we adopt the joint distribution function (Catelan \&
Theuns 1996),
\begin{eqnarray}
P(\lambda|\nu) & = & \frac{0.68}{\lambda}
\left[1+0.02\left(\frac{\lambda}{\lambda_0(\nu)}\right)^4\right] \times 
\nonumber \\
& & \exp\left[-\frac{\log ^2(\lambda/\lambda_0(\nu))}{0.98}\right],
\end{eqnarray}
where $\lambda_0(\nu)=0.11 \nu^{-1.1}$ and
$\nu=\nu(z)=\delta_c(z)/\sigma(R,z)$.  Since, at a fixed epoch, there is not a
one-to-one relationship between $\nu$ (defined for the spatially averaged
overdensity field) and mass, owing to the distribution of halo shapes,
equation (12) effectively corresponds to an average of $\nu$ over all shapes
of a given mass.\footnote{We point out that the binding energy used to
calculate $\lambda$ is derived using a spherical-accretion model, but that for
the lower, more irregular peaks, this is likely only a lower limit to the
actual binding energy, resulting in an underestimate of $\lambda$. Combined
with the arguments above from Catelan \& Theuns (1996), one can push the
predicted spin values for spirals (if these are indeed associated with
lower-$\nu$ peaks, as opposed to ellipticals, which are associated with higher
peaks) closer to the observed values, such that one needn't rely as heavily on
dissipation to spin up spirals.}

\subsection{Chemical Evolution}

HJ99 assumed the stellar populations to have constant metallicity fixed at the
solar value for all time. More realistic models should, of course, account for
chemical evolution.  To study the effect of chemical evolution on the TF
relation, we assume that galaxies are (chemically) closed boxes, for which
analytic results for the evolution of metallicity exist (e.g., Pagel
1997). This assumption seems to be justified by the detailed hydrodynamical
computations of a multi-phase ISM by Mac Low \& Ferrara (1999).  In order to
relax the instantaneous recycling approximation we adopt a value of 0.03 for
the yield in order to account for the delayed production of Fe by Type-Ia
supernovae. We also set the zero-point redshift dependence so as to comply
with observations of damped Lyman-$\alpha$ systems, which suggest that the ISM
is already enriched up to a certain metallicity at a given redshift. Adopting
the relation derived by Pettini \etal (1999; see Figure 8 therein), we have
that the metallicity, $Z$, evolves with redshift, $z$, as
\begin{equation}
Z(z,z_f)=0.03 Z_{\odot} \ln\left(\frac{1}{f_g(z,z_f)}\right) + 0.28 Z_{\odot} 10^{-0.25z},
\end{equation}
where the solar metallicity $Z_{\odot}=0.02$ and $f_g(z,z_f)$ is gas fraction
of the disk.  It will be shown that chemical evolution will be an important
factor in reducing the predicted scatter in the TF relation.

\section{Results}

We first study the respective impacts of the various ingredients which have
entered into our galaxy-formation model.  In Figures \ref{fig2} to
\ref{fig12}, we illustrate examples of how our predictions for the TF relation
in the $B$, $R$, $I$, and $K$ bands change as the model inputs are varied.
These results will allow the reader to infer how the predictions of any of our
models, or any other models that appear in the literature, would change with
different input physics or assumptions.  In each graph, the solid line
delineates the power-law fit to the TF-relation prediction (along with
1-$\sigma$ errors given by the dashed lines, and denoted in each plot by
`$\sigma$'), and the four scattered-dot curves trace the spread in the
predicted TF relation for four fixed masses ($10^{10}$, $10^{11}$, $10^{12}$,
and $10^{13} M_{\odot}$), using $\sim 150$ points each.

Overplotted on each graph are open symbols representing data from Tully \etal
(1998). These data, comprised of spiral galaxies in the loose clusters of Ursa
Major and Pisces, represent one of the most carefully defined TF samples. The
inclination-corrected FWHM of the lines has been converted to $W^i_R$, which
approximates to $2 V_c$, and the data have been corrected for dust
extinction. The scatter in the $B$, $R$, $I$, and $K$ bands is 0.50, 0.41,
0.40, and 0.41 mag, respectively. While many authors compare their models to
TF data by defining a ``fundamental'' TF relation derived from the particular
data set in question, we argue that, given the observational selection effects
inherent in defining such samples, there is no particular significance or
accuracy associated with any one determination of the slope and normalization
of the TF relation. Instead, we examine how well these data fit to our model,
in a $\chi^2$ sense, using the measured scatter as an estimate of the error
associated with each point.  For each plot we calculate the resulting $\chi^2$
and denote by `$p$' the probability of obtaining this large a value of
$\chi^2$ given that the model is correct; values of $p$ below $10^{-3}$ are
listed as zero.  In addition, since the data have excluded spirals which show
evidence of merger activity or disruption in the form of starbursts, we
exclude from our model galaxies those having $B-R < 0.3$.

\subsection{Formation-Redshift Distribution}

\subsubsection{S Distribution}

\begin{figure*}
\centerline{
\psfig{figure=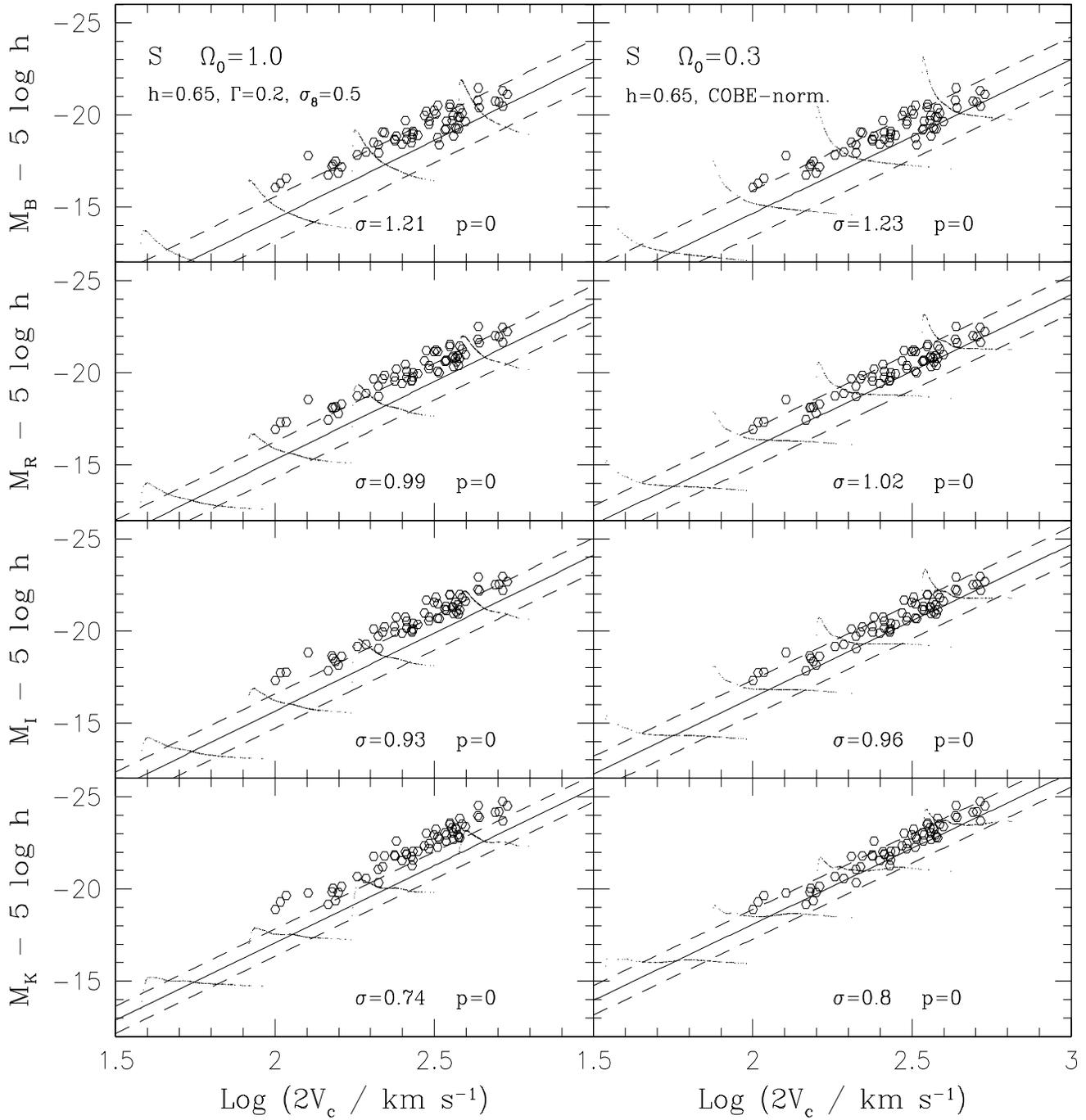,height=20cm,width=18.5cm,angle=0}}
\caption{\label{fig2} S-distribution predictions. The left panels show the
predictions for the present-day TF relation in $B$, $R$, $I$, and $K$ for the
S distribution assuming an EdS universe with $h=0.65$ and a CDM power spectrum
with $\Gamma=0.2$ and $\sigma_8=0.5$. The right panels display the results for
a COBE-normalized $\Lambda$CDM model with $\Omega_0=0.3$, $h=0.65$, and
$\Gamma=\Omega_0 h$. Both assume a Salpeter IMF, a constant spin parameter of
$\lambda=0.05$, and constant metallicity fixed at the solar value.  The model
predictions for halos of mass $10^{10}$, $10^{11}$, $10^{12}$, and
$10^{13}M_\odot$ are indicated by the four scatter-point curves in each panel,
and the solid lines are the fits to these predictions, with 1-$\sigma$ errors
indicated by the dashed lines. The open symbols are data from Tully \etal
(1998). Note the EdS model, in particular, underpredicts the luminosity in all
bands.}
\end{figure*}

As discussed above, the distribution of formation redshifts should lead to
some scatter about the TF relation, and Figure \ref{fig2} illustrates this
effect using the S distribution.  We show results for two cosmogonies to
indicate the dependence of the results on some of the cosmological parameters.
The left-hand panels of Figure \ref{fig2} depict the results for the S
distribution of $z_f$ in an EdS universe, while the right-hand panels show the
results for a flat, $\Lambda$-dominated universe. Here, and unless otherwise
indicated, we shall take EdS models to have $h=0.65$, a power-spectrum shape
parameter\footnote{Since values of $\Omega_0 h \ga 0.3$ already seem to be
ruled out (e.g., Peacock \& Dodds 1994), $\Gamma$ in the case of EdS models is
defined simply as a fitting parameter in the CDM transfer function. In the
case of $\Lambda$CDM models, we formally take $\Gamma = \Omega_0 h$.}
$\Gamma=0.2$, and an rms density contrast fluctuation over 8$h^{-1}$ Mpc
spheres of $\sigma_8=0.5$.  For $\Lambda$CDM models, we assume $\Omega_0=0.3$,
$h=0.65$, a COBE-normalized power spectrum (Bunn \& White 1997) with
$\Gamma=\Omega_0 h=0.195$, and for all cosmogonies we assume $\lambda=0.05$, a
fixed metallicity given by the solar value ($Z=Z_{\odot}$), and a Salpeter
IMF, $d\log{N_*}/d\log{M_*}=-\alpha$ with $\alpha={1.35}$, all unless
otherwise indicated.

To understand the results, we begin by examining the predictions for the
$K$-band TF relation in the $\Lambda$CDM model. Here we find that the spread
in $z_f$ translates directly into a spread in $V_c$ (with earlier formation
implying higher circular velocity) which widens for lower masses, since these
can form over a broader range of redshifts stretching back to higher $z$. This
broadening of the TF with lower $V_c$ arises generically from a realistic
formation-distribution, and it is observed in most TF samples.  The predicted
$M_K$, however, varies very little for galaxies of a given mass, since this
wavelength primarily traces the older stellar population.  In general for our
models, the distribution in $z_f$ is found to be the primary source of scatter
in the TF relation.

As we go to bluer bands, we find that for a given mass, younger galaxies
become brighter, increasing the spread in the TF relation. This is a simple
consequence of equation (8) in HJ99, which states that the SFR is a rapidly
declining function of age. There is a competing effect, however, from the fact
that galaxies that are too young will not have had enough time to convert much
gas into stars, and so will become dimmer. For this reason, our plots often
outline a peak magnitude for galaxies of a given mass, in those cases where
the $z_f$ distribution produces a large fraction of very young galaxies. Such
galaxies would likely exhibit the properties of starbursts and be excluded
from carefully-defined TF samples. Indeed, model galaxies removed by our color
selection criterion ($B-R<0.3$) tend to be those on the younger side of these
peaks.

Overall, these S distribution predictions do not fit the TF data in any
waveband, as the models are too faint in every case, and the predicted scatter
here is too large by about 0.7 mag in $B$ and 0.3 mag in $K$. The problem owes
largely to the fact that the S distribution has a substantial high-redshift
tail producing older, fainter galaxies with high $V_c$. Thus, if we assume
star formation proceeds according to a Schmidt law, TF observations imply that
{\em the majority of present-day disks could not have formed much beyond a
redshift of} 2--3.

\subsubsection{LC Distribution}

\begin{figure*}
\centerline{
\psfig{figure=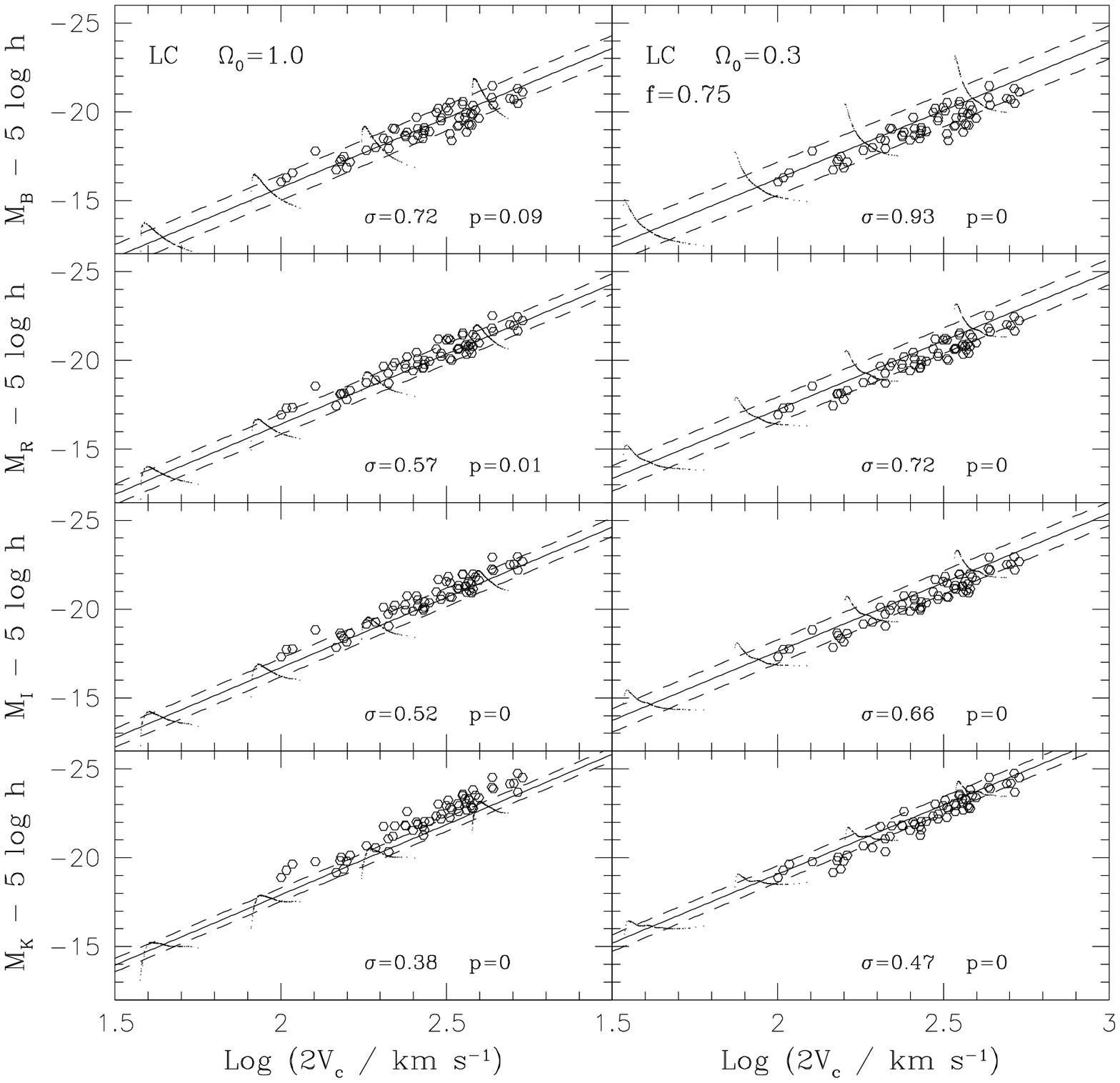,height=21cm,width=18.5cm,angle=0}}
\caption{\label{fig3} LC-distribution predictions. 
Same as Figure \ref{fig2}, but using the LC distribution with $f=0.75$.}
\end{figure*}

Figure \ref{fig3} depicts corresponding results for the same parameter
choices, but using the LC distribution with $f=0.75$. This distribution has
both a lower peak value of $z_f$ and a smaller FWHM than the S distribution
(see Figure \ref{fig1}), producing significantly less scatter in the predicted
TF relations and a larger fraction of younger galaxies, as compared with
Figure \ref{fig2}.  Despite the smaller scatter, these models still do not fit
the observed TF relation in the various bands. Note that the EdS model here
actually yields excellent agreement in $B$, but predicts galaxies which are
too faint in $I$ and $K$.

\begin{figure*}
\centerline{
\psfig{figure=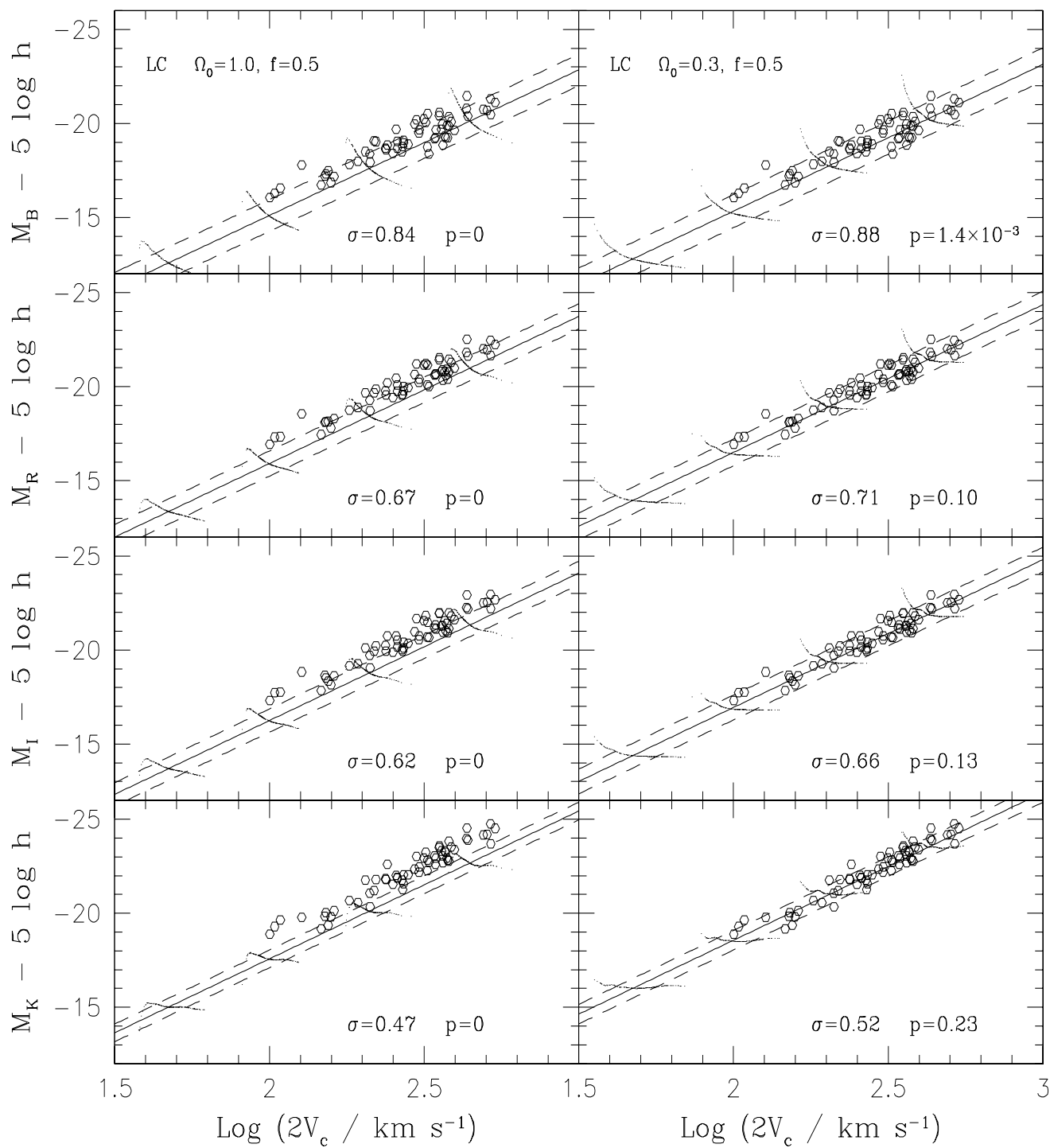,height=21cm,width=18.5cm,angle=0}}
\caption{\label{fig4} Same as Figure \ref{fig3}, but using $f=0.5$. Note how
the $\Lambda$ model yields excellent agreement in $K$, $R$, and $I$, but
provides a relatively poorer fit in $B$.}
\end{figure*}

\begin{figure*}
\centerline{
\psfig{figure=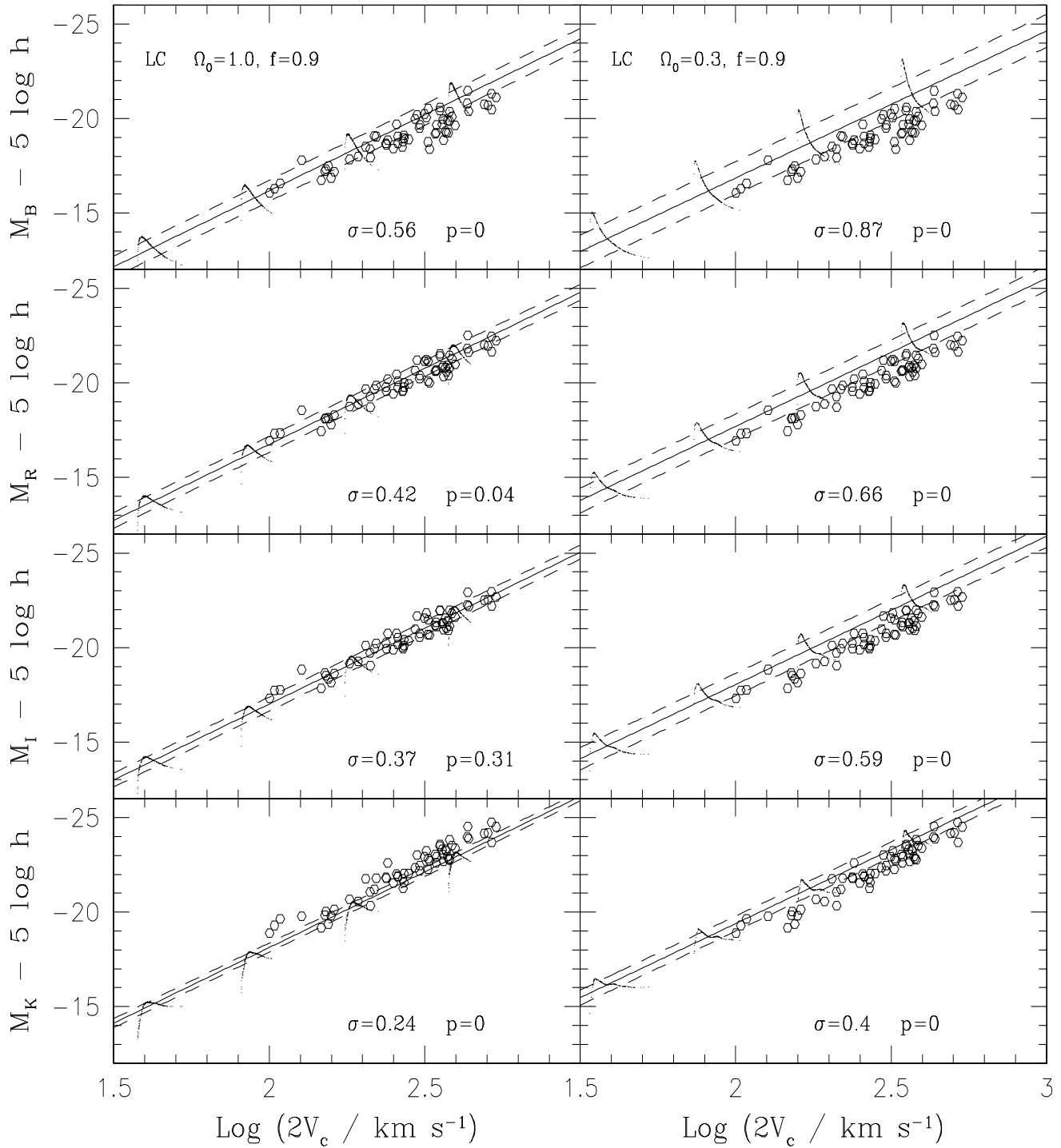,height=21cm,width=18.5cm,angle=0}}
\caption{\label{fig5} Same as Figure \ref{fig3}, but using $f=0.9$. In this
case, all disks form very recently and many fall on the younger side of the
magnitude peak.  These effects combine to yield very little TF scatter for the
few relatively older galaxies which fall red-ward of our color selection
criterion.  The overall agreement however, is poor.}
\end{figure*}

In Figures \ref{fig4} and \ref{fig5} we examine LC distributions with values
of $f=0.5$ and $f=0.9$, respectively.  Lacey \& Cole (1993, 1994) found that
the statistics of the time at which present-day halos accreted half their mass
(i.e., $f=0.5$) provides an excellent fit to the results of $N$-body
simulations. Here we see that the $f=0.5$ $\Lambda$CDM model provides an
excellent fit to the $K$-, $R$-, and $I$-band TF relations but predicts
galaxies which are too faint in $B$, thus highlighting the importance of using
multi-wavelength constraints. This model also produces TF scatter which is
about 0.1 to 0.4 mag too large, going from $K$ to $B$. The EdS predictions
here have less scatter, but are again too faint in every case. In general, we
find that LC distributions with plausible values of $f$ fare better than S
distributions at matching TF data, suggesting disk formation in the range
$0<z<2$ (see below).

The $f=0.9$ models have a $z_f$ distribution which produces galaxies only in a
narrow interval centered at low redshift (see Figure \ref{fig1}).  This
results in extremely small TF scatter in part because effectively no galaxies
form above a relatively low maximum redshift of $z \sim 1$, but also because
many of these recently formed galaxies fall on the young side of the magnitude
peak and are excluded by our color selection. In effect, this $z_f$
distribution has ``pushed'' galaxies over to the faint side of the magnitude
peak, where their colors resemble those of starbursts, and the low scatter
owes to the small number of galaxies ``older'' than the peak, particularly in
$K$. Thus, while the intrinsic TF scatter of all galaxies in this model is
high, the observed scatter, based on observational selection, can be quite
low, and this effect is important when considering recently formed
populations.

Irrespective of the scatter, however, the $f=0.9$ models fail to fit the TF
slope and normalization. For the reasonable cosmological-parameter values in
the right panels, the predicted galaxies are very young and thus excessively
bright. Thus, TF constraints also imply that {\em the majority of disk
formation cannot have happened at} $z < 1$.  Of course, from a physical
standpoint, such large values of $f$ are anyway unreasonable. Note that even
with $f=0.9$, the EdS model still produces galaxies which are too faint in
$K$, despite yielding excellent agreement in $I$.

\begin{figure*}
\centerline{
\psfig{figure=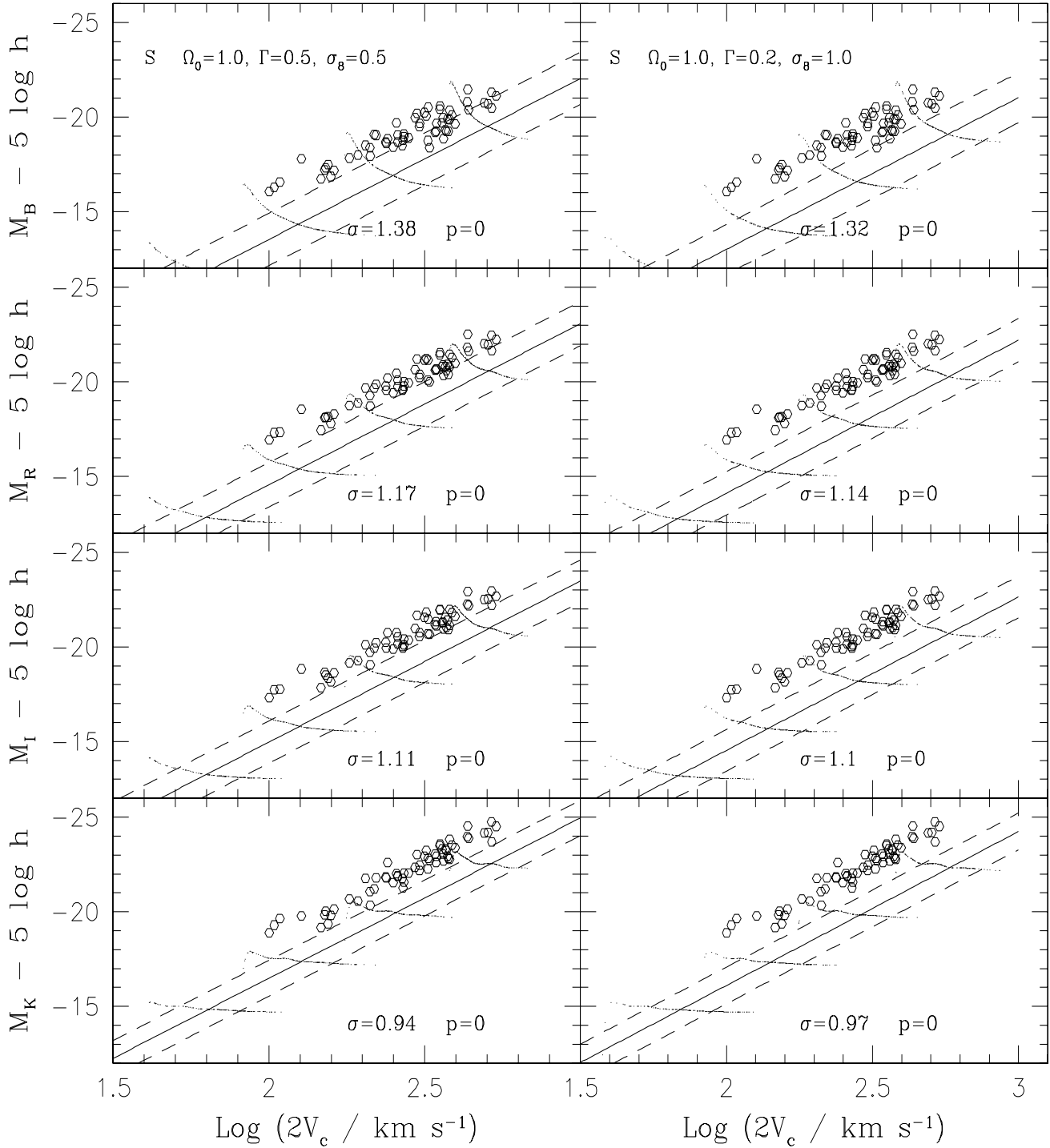,height=21cm,width=18.5cm,angle=0}}
\caption{\label{fig6} Dependence on the power spectrum shape and amplitude.
TF predictions for the S distribution in an EdS universe for $\Gamma=0.5$ and
$\sigma_8=0.5$ (left panels), and for $\Gamma=0.2$ and $\sigma_8=1.0$ (right
panels). Comparing with Figure \ref{fig2}, we find that high $\Gamma$ and high
$\sigma_8$ both lead to earlier collapse, producing fainter disks with a much
larger TF scatter.}
\end{figure*}

\begin{figure*}
\centerline{
\psfig{figure=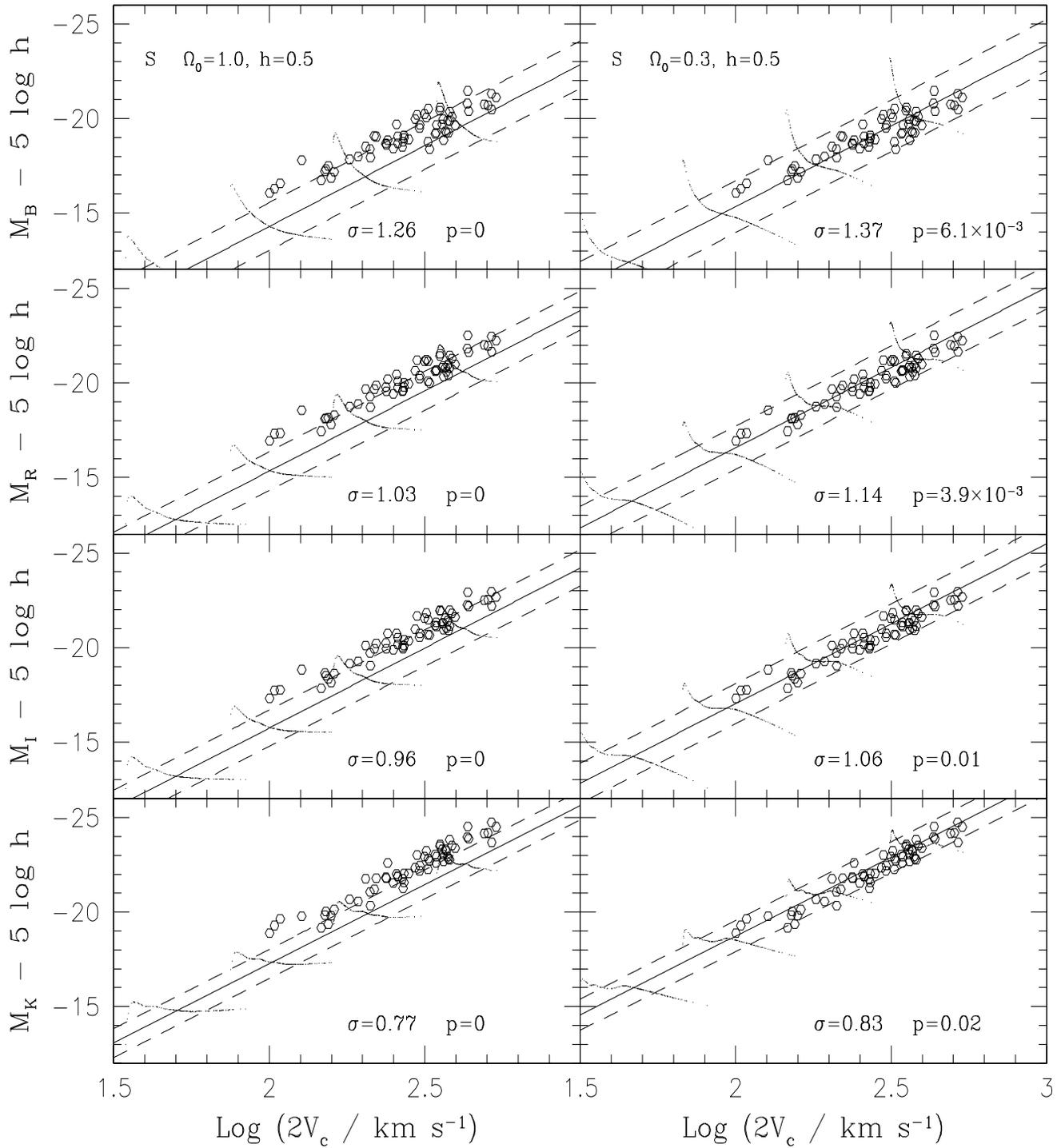,height=21cm,width=18.5cm,angle=0}}
\caption{\label{fig7} Dependence on the Hubble constant.  Same as Figure
\ref{fig2}, but using $h=0.5$. The resulting changes in baryon fraction and
age lead to a brightening of the TF relation, particularly for low
$\Omega_0$.}
\end{figure*} 

\begin{figure*}
\centerline{
\psfig{figure=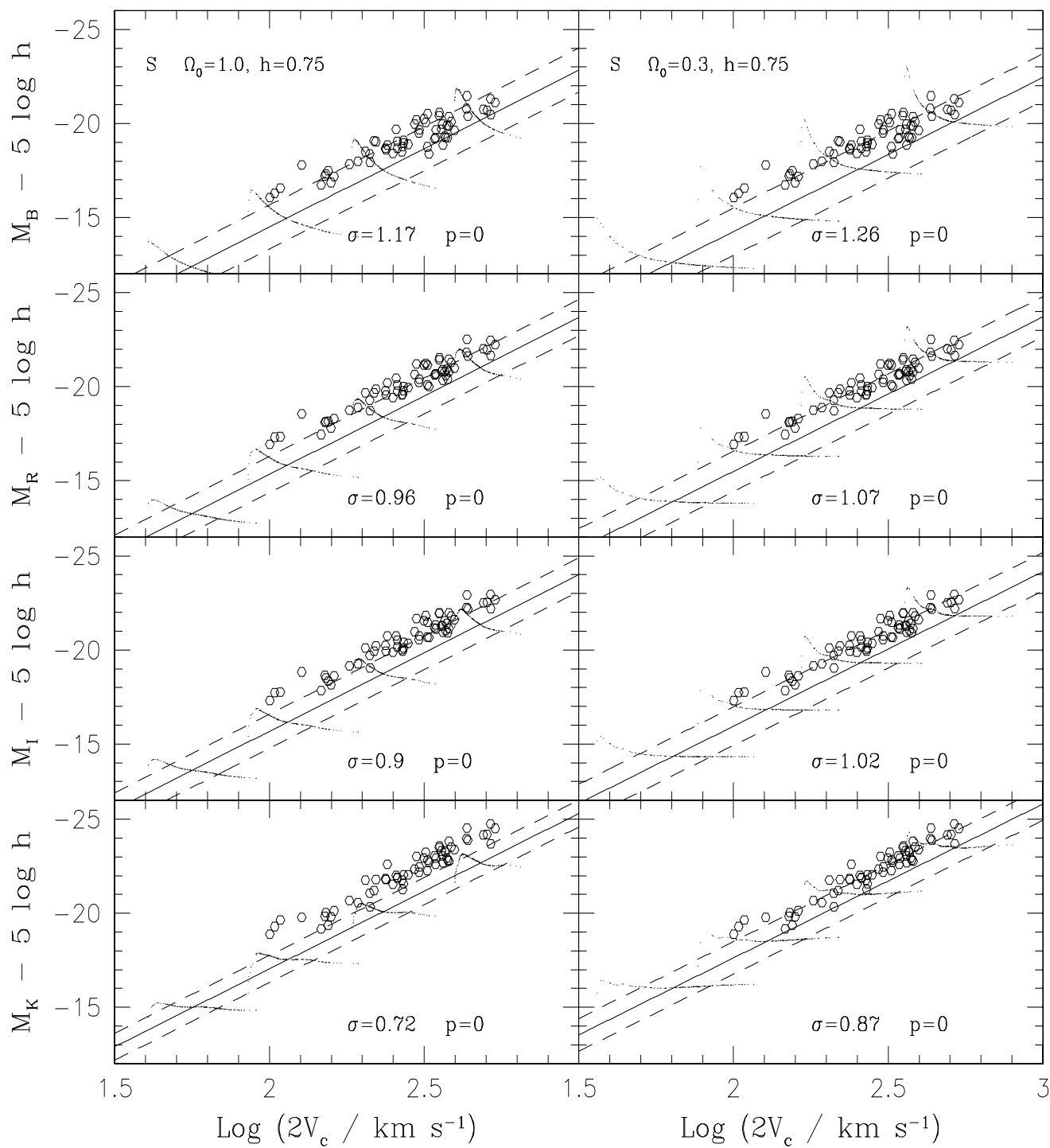,height=21cm,width=18.5cm,angle=0}}
\caption{\label{fig8} Same as Figure \ref{fig2}, but using $h=0.75$.}
\end{figure*} 

\begin{figure*}
\centerline{
\psfig{figure=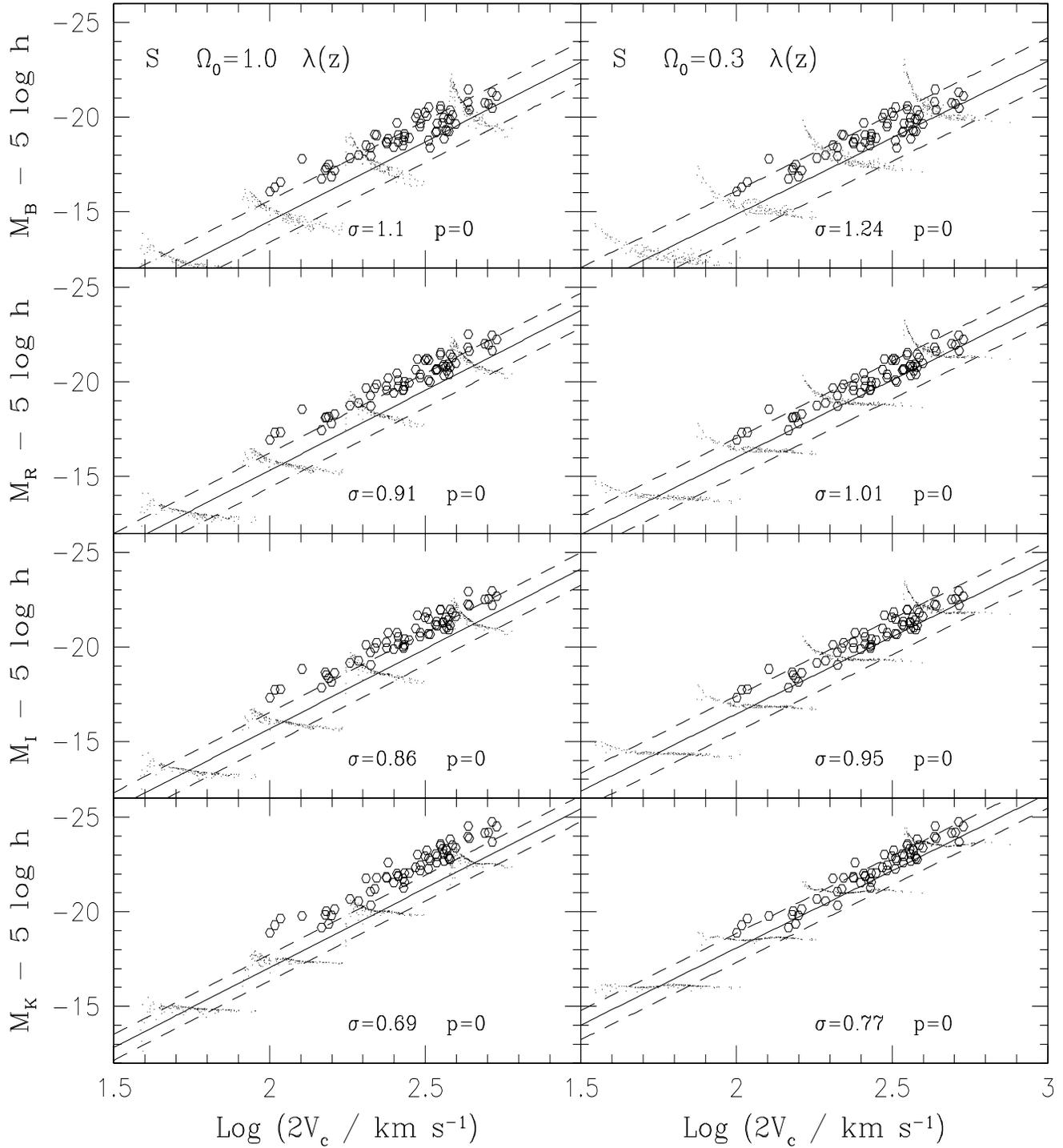,height=21cm,width=18.5cm,angle=0}}
\caption{\label{fig9} Effect of joint distribution in spin and peak
height. Same as Figure \ref{fig2}, but now incorporating both the predicted
shape of the spin-parameter distribution and the anti-correlation between spin
and peak height. These results differ little from those of Figure \ref{fig2};
the joint distribution in $\lambda$ and $\nu$ scatters the predictions, by a
small amount, along the TF relation itself, tightening the TF relation in some
cases.}
\end{figure*}

\begin{figure*}
\centerline{
\psfig{figure=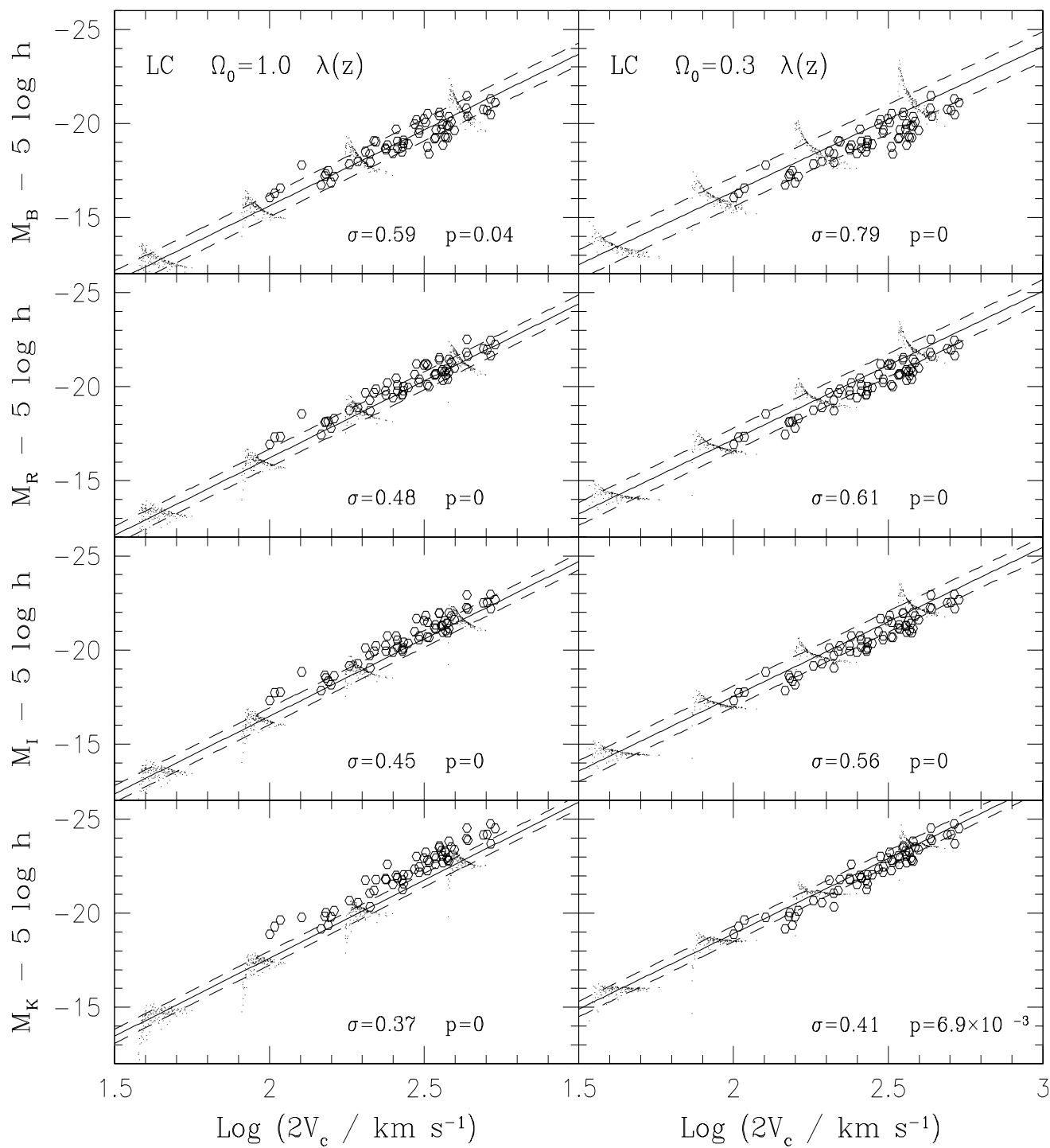,height=21cm,width=18.5cm,angle=0}}
\caption{\label{fig10} Same as Figure \ref{fig3}, but now incorporating the
joint PDF in $\lambda$ and $\nu$ into the LC results.}
\end{figure*}

\begin{figure*}
\centerline{
\psfig{figure=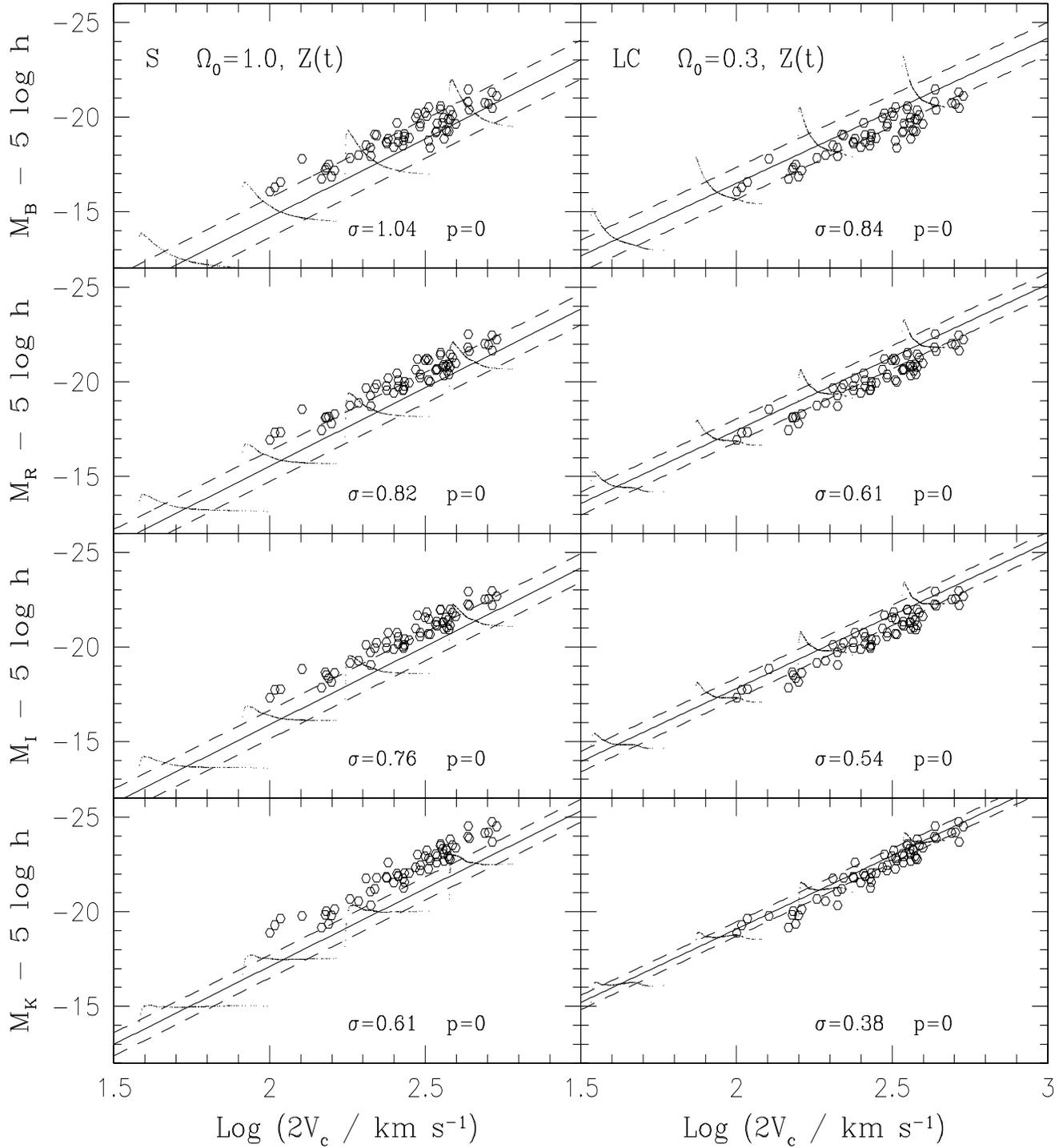,height=21cm,width=18.5cm,angle=0}}
\caption{\label{fig11} Effect of chemical evolution. TF predictions for an
$\Omega_0=1$ S distribution (left panels) and an $f=0.75$ LC distribution with
$\Omega_0=0.3$ (right panels), both with evolving metallicity.  Note the
smaller TF scatter as compared to the corresponding results in Figures
\ref{fig2} and \ref{fig3}.}
\end{figure*}

\begin{figure*}
\centerline{
\psfig{figure=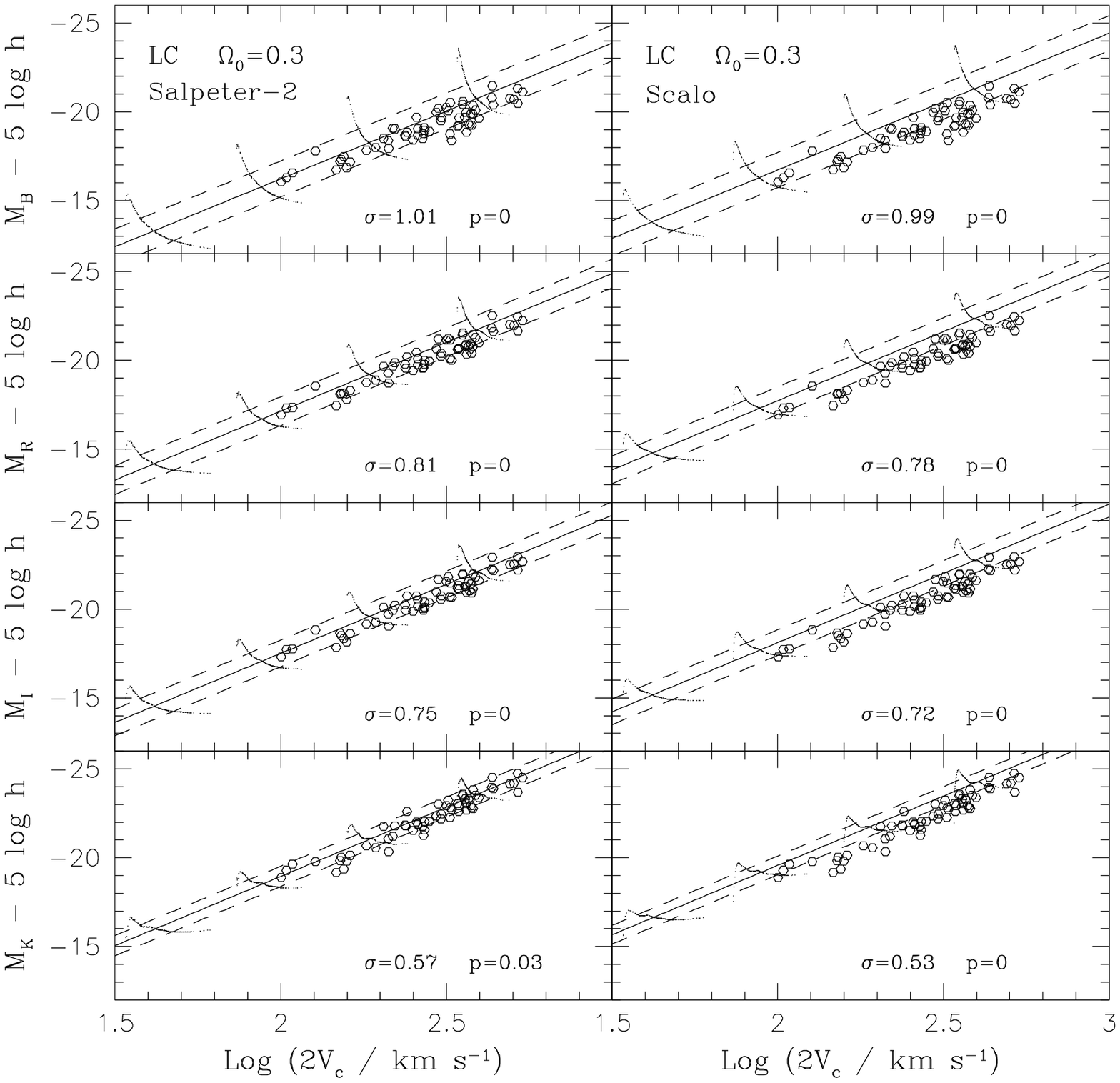,height=21cm,width=18.5cm,angle=0}}
\caption{\label{fig12} Dependence on the IMF. TF predictions for an $f=0.75$
LC distribution with $\Omega_0=0.3$, assuming a Salpeter IMF with index
$\alpha=0.95$ (left panels), and a Scalo IMF (right panels).  These IMFs
produce relatively more massive stars than the $\alpha=1.35$ Salpeter IMF
investigated previously, and thus brighter TF predictions than those in the
right panels of Figure \ref{fig3}.}
\end{figure*}

\subsection{Dependence on Cosmogony}

\subsubsection{Power-Spectrum Shape and Amplitude}

In Figure \ref{fig6} we explore the impact of individually changing the shape
and normalization of the power spectrum, using the S distribution in an EdS
universe.  In the left (right) panels we adopt $\Gamma=0.5$ ($\Gamma=0.2$) and
$\sigma_8=0.5$ ($\sigma_8=1.0$).  In both cases, the resulting galaxies are
several magnitudes too faint and produce dramatically larger scatter in their
TF relation.  In the case of higher $\sigma_8$, halos of a given mass will be
smaller, and correspond to rarer peaks. These collapse at earlier times,
leading to larger $V_c$ and older, dimmer galaxies.  Extending the range of
collapse redshifts back to higher $z$ also substantially increases the
scatter.  Similarly, in these CDM models, increasing $\Gamma$ shifts more
power to smaller scales ($R \la 10$ Mpc), leading to the same result.  Though
we do not explicitly explore variations with $n$, it can be inferred that
values of $n > 1$, which also tilt power towards smaller scales, will have a
similar effect.  Conversely, values below unity might lead to better agreement
with the data, but there are already strong constraints on the tilt (e.g.,
Kamionkowski \& Buchalter 2000).

\subsubsection{Hubble Constant and Matter Density}

Given the uncertainties in our present understanding of galaxy formation, it
makes little sense to attempt to constrain cosmological parameters using TF
data.  Rather, with these parameter values dictated by other, more direct
tests, we should employ the TF relation to gain insight into these details.
Variations in $\Omega_0$ and $h$ affect almost every aspect of the model, such
as the disk mass fraction, galaxy ages and circular velocities, star-formation
rates, and the power-spectrum shape. In Figures \ref{fig7} and \ref{fig8} we
plot S distributions with $h=0.5$ and $h=0.75$, respectively.  Little change
is seen in the EdS cases, while for low $\Omega_0$ the net result appears to
be that lower (higher) values of $h$ produce galaxies which are brighter
(fainter).  Changes in $h$ do not, however, strongly affect the scatter.  Note
that the $h=0.5$ $\Lambda$CDM model provides reasonable agreement in $K$ and
$I$, but is not as successful in $B$ or $R$.  The dependence of the models on
$\Omega_0$ can be gleaned from comparing the left and right panels of the
various Figures. In general, for a fixed halo mass, lower values of $\Omega_0$
will yield a larger disk mass and thus brighter luminosity. We have seen
repeatedly that high values of $\Omega_0$ invariably predict galaxies which
are excessively faint as compared to observations.

In order to gain broader insight into the dependence of our results on the
cosmological parameters, we generated data for the $\Omega_0=1$ case with
$\Gamma=0.2$ and $h=0.45$--0.75, and for $\Omega_0=0.2$--0.5 cases with
$\Gamma=\Omega_0 h$ and $h=0.55$--0.75, both for the S distribution and for LC
distributions with $f=0.5$ and 0.75. All were COBE normalized. Only a handful
of these models produced values of $p>0.001$ in any band, and those which did
invariably had values of $\Omega_0 h \sim 0.2$, as required by current
measurements. EdS models generally produce galaxies which are too faint,
unless unacceptably low values of $h$ are adopted, and produce too large a
scatter, unless unacceptably low values of $\sigma_8$ are adopted.  The best
fitting models, both in terms of $p$ and scatter, were LC distributions with
COBE normalization, $\Omega_0 \sim 0.3$, and $h \sim 0.65$, similar to the
best-fitting values obtained in earlier work (EL96; van den Bosch 2000) and in
agreement with current estimates.  As mentioned above, the S distribution
predictions are generally too faint, due to the significant fraction of older
galaxies arising from the high-redshift tail.

\subsection{Spin-Parameter Distribution}

We now turn to the issue of the spin-parameter distribution. Figures
\ref{fig9} and \ref{fig10} correspond to the S and LC distributions from
Figures \ref{fig2} and \ref{fig3}, but with the joint probability distribution
in $\lambda$ and $\nu$ now taken into account.  In the S distributions, this
produces virtually no effect in $K$, as expected, but does broaden the
distribution of points at each mass in the bluer bands. In every case,
however, the TF scatter, is roughly similar to, or less than, that in Figure
\ref{fig2}; the broadening has aligned along the TF relation itself. This
arises from the fact that for a given mass, higher peaks (which collapse
earlier and have higher $V_c$) will have smaller spins leading to higher gas
surface density and thus higher luminosities.  A similar result is found in
the LC case, though here the $z_f$ distribution is weighted toward more recent
epochs, exaggerating this effect and leading to a larger systematic reduction
in the scatter, once our color selection criterion is enforced. Thus, in
practice, accounting for the joint probability distribution in $\lambda$ and
$\nu$ can reduce the TF scatter by about 0.15 mag in $B$ to 0.05 mag in $K$,
for plausible cosmogonies. Otherwise, the spin distribution does not have a
great impact on the TF relation. As a corollary to this, we find that
high-surface-brightness and low-surface-brightness disks, as distinguished by
different $\lambda$ or $R_d$ in our model (see Figure \ref{fig14}), are
predicted to lie on the same TF relation, as is in fact observed (Tully \etal
1998).

\subsection{Chemical Evolution and the IMF}

The role of chemical evolution is investigated in Figure \ref{fig11} for the
$\Omega_0=1$ S distribution and the $\Lambda$CDM LC distribution. Compared
with their constant, solar-metallicity counterparts in Figures \ref{fig2} and
\ref{fig3}, these models have TF scatters which are systematically smaller by
about 0.1 mag in all wavebands. This simply results from the fact that
relative to constant metallicity systems, older disks (with higher $V_c$) will
continually be forming populations of higher metallicity, and thus have a
higher integrated luminosity.  Conversely, younger systems will be relatively
dimmer, so that models with an evolving metallicity content will naturally
align more tightly along a luminosity-circular velocity relationship than
those with constant $Z$.

In Figure \ref{fig12} we examine the impact of changing the assumed IMF. The
left panels depict a Salpeter IMF with $\alpha={0.95}$ (denoted as
Salpeter-2), while the right panels represent a Scalo IMF (Scalo, 1986), each
for the $\Lambda$CDM LC distribution. These IMFs both produce relatively more
high-mass stars, thus yielding predictions which are excessively bright in the
bluer bands. Note that the Salpeter-2 model matches reasonably in $K$, but not
in any other band. This is yet another example of the importance of using
multi-wavelength constraints to discriminate between models which may appear
to fit in a single band.

\begin{figure*}
\centerline{
\psfig{figure=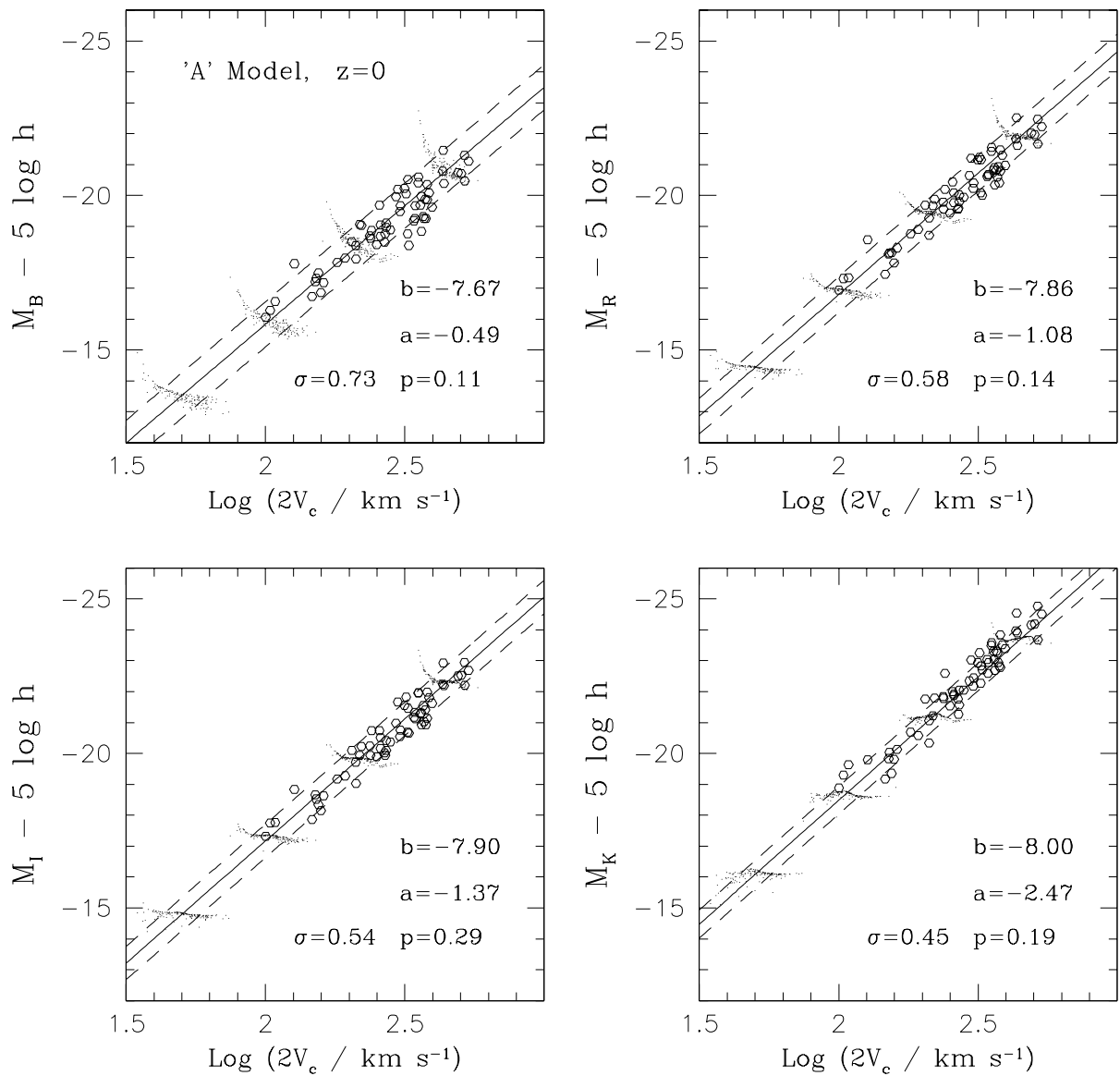,height=17cm,angle=0}}
\caption{\label{fig13} Agreement of the A model.  Present-day TF predictions
for the A model in $B$, $R$, $I$, and $K$ for a COBE-normalized $f=0.5$ LC
distribution with $\Omega_0=0.3$, $h=0.68$, a Salpeter IMF, including the
joint distribution in $\lambda$ and $\nu$ as well as chemical evolution.  This
model, denoted as the `A' model, produces excellent agreement in all bands
($p>0.10$) with roughly the correct amount of scatter ($\sigma \sim 0.4$--0.5).
}
\end{figure*}

\begin{figure*}
\centerline{
\psfig{figure=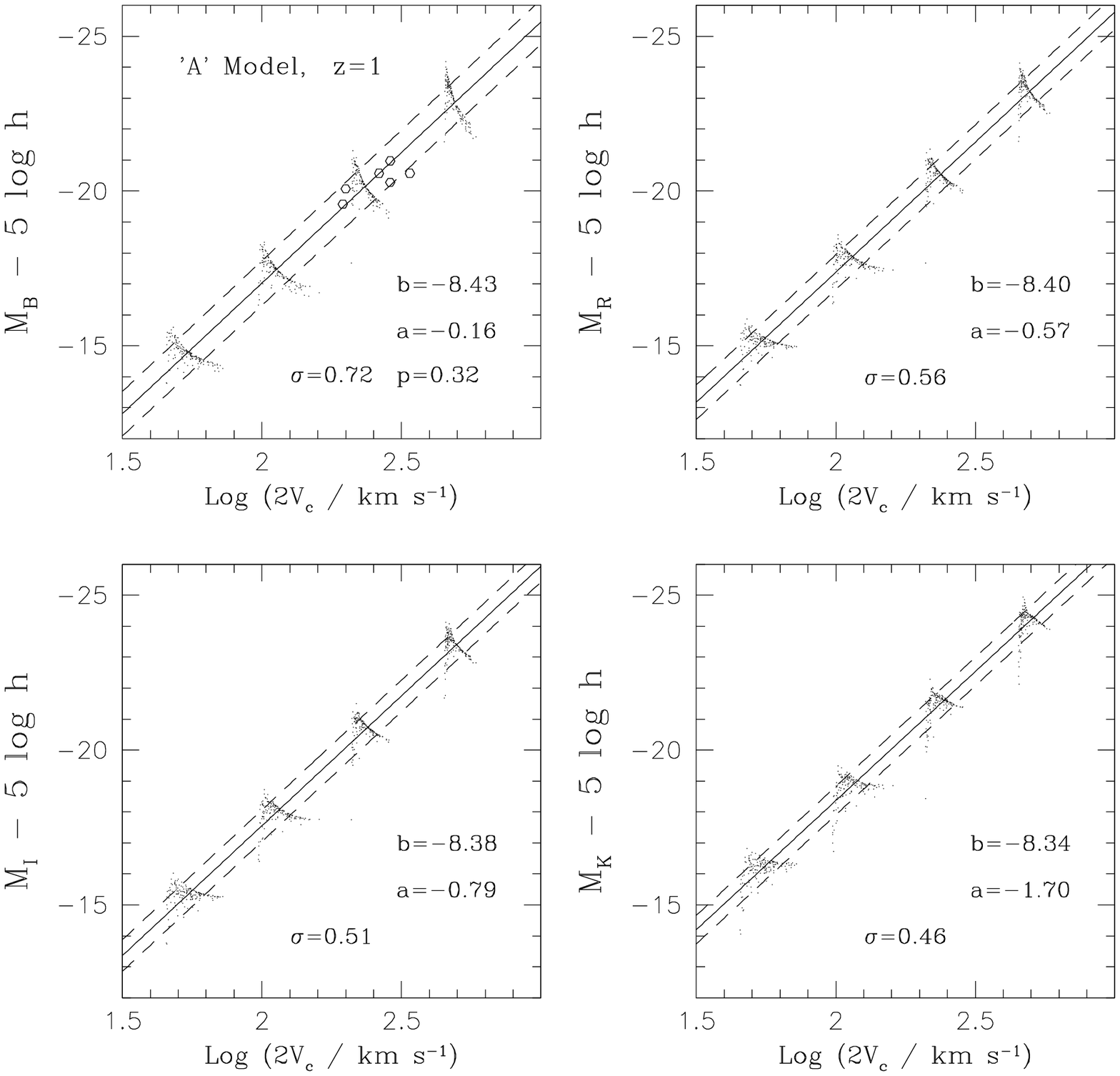,height=17cm,angle=0}}
\caption{\label{fig14} High-redshift predictions.  TF prediction of the A
model for $z=1$, along with 6 $B$-band data point from Vogt \etal (1996) for
spirals in the range $0.5<z<1$.  The $z=1$ prediction have steeper slopes
(given by $b$) and lower zero-points (given by $a$), than the counterpart
$z=0$ predictions, resulting in a brighter $B$-band TF relation, but roughly a
roughly similar relation in $K$.}
\end{figure*}

\section{A Model That Works}

\subsection{The TF Relation at z=0 and z=1}

As an example of what can be accomplished with the theoretical framework
presented here, we illustrate specifically the results for a COBE-normalized
$\Lambda$CDM LC distribution with $f=0.5$, $\Omega_0=0.3$, $h=0.68$, a
Salpeter IMF, and with the joint distribution in $\lambda$ and $\nu$, as well
as chemical evolution. We refer to this particular model as the `A' model.
While this is not necessarily the best-fitting model over our entire parameter
space, it does yield remarkable agreement with the data, as shown in Figure
\ref{fig13}, and these model parameters fall nicely in line with current
observational constraints. In addition to $\sigma$ and $p$, the plots list the
predicted normalizations and slopes [given by $a$ and $b$ in equation
(\ref{tf})].  The A model appears to fit the slope, normalization, and scatter
of the TF relation reasonably well in all bands. The predicted scatter in $B$
remains about 0.2 mag higher than observed, but there are several reasons one
might expect this, given our model assumptions.  In the next section, we show
that features such as strong supernova feedback or a flatter density profile
towards the core are not necessary to successfully predict many of the global
spiral-galaxy properties investigated here. Thus, these features would tend
only to produce small changes in the amount of gas available for star
formation, so that their effects would be largely confined to $B$. Presumably,
however, their inclusion would result in a slight dimming and tightening the
$B$-band predictions, bringing them more in line with the observations.

Figure \ref{fig14} shows the predictions of the A model for the TF relation at
$z=1$, together with 6 data points from Vogt \etal (1997) for spirals in the
range $0.5<z<1.0$. These data have been corrected for dust.  At $z=1$ our
model predicts larger values of $a$ (i.e., lower zero-points) and steeper
slopes for the TF relation in every band. This results in the TF relation
becoming about 1 mag brighter at 50 km s$^{-1}$, and about 2 mag brighter at
500 km s$^{-1}$ in the $B$-band, but conspires to produce very little change
at these scales in the $K$ band. Moreover, the scatter in the TF relation in
all bands at $z=1$ is predicted to be roughly the same as that at $z=0$,
though we do not impose any color-selection criterion in the high-redshift
case, since it is not as clear what the observational selections are. Though
the $B$-band data shown are comprised of only 6 points, they have an intrinsic
scatter consistent with that of the low-$z$ data, and are fit by our model
extremely well ($p=0.32$), though, as in the $z=0$ case, the predicted scatter
is about 0.2 mag too large.

\begin{figure}
\centerline{
\psfig{figure=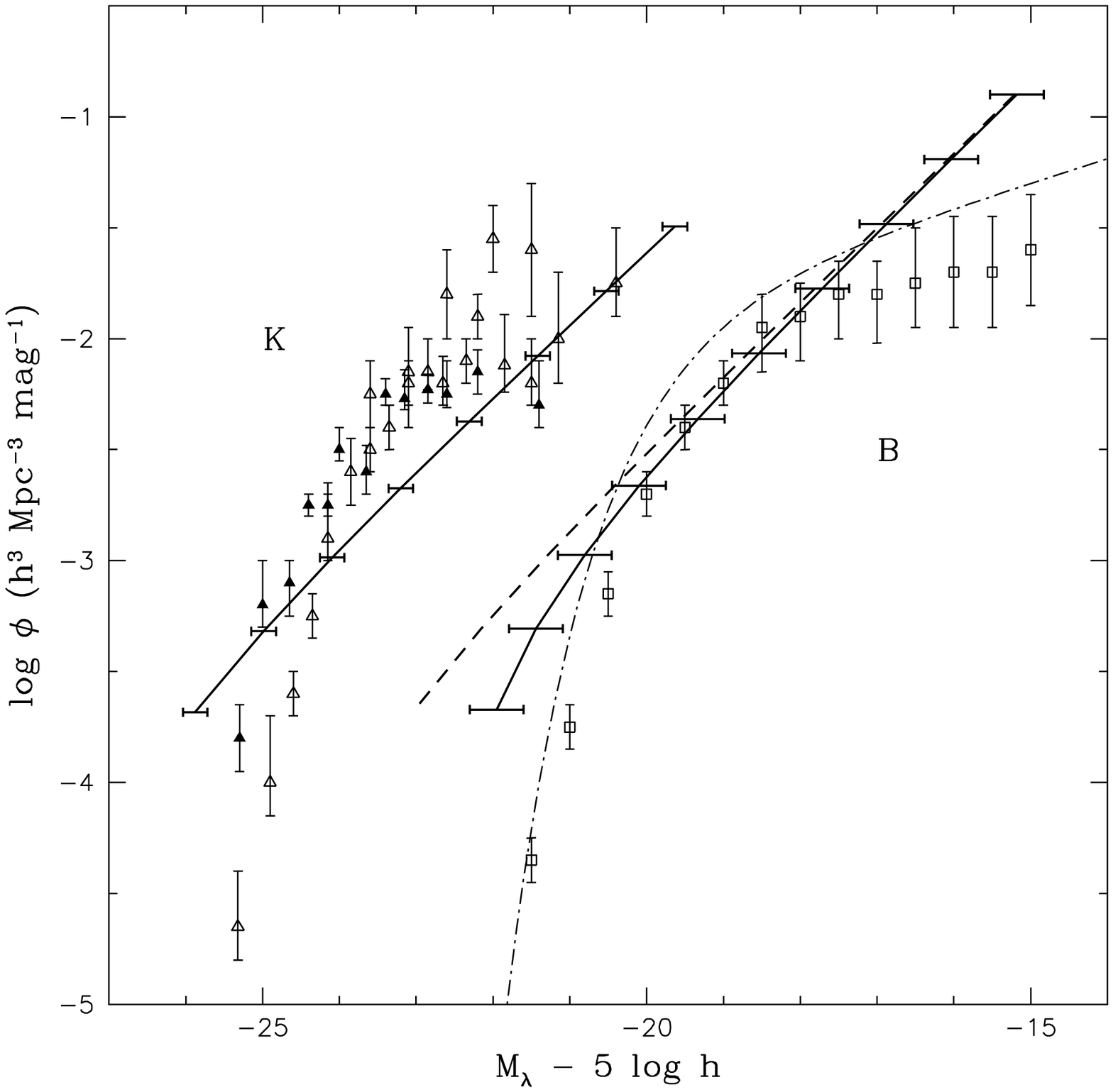,height=8cm,angle=0}}
\caption{\label{fig15} The luminosity function.  $B$ and $K$ LF predictions of
the A model (solid lines) together with 1-$\sigma$ errors (horizontal error
bars). The dashed line is the $B$-band prediction before correcting for dust.
The dot-dashed line is the $B$-band fit to the 2dF Survey data (Folkes \etal
1999), and the open squares are data from Zucca \etal (1997).  The $K$-band
data are taken from Gardner \etal (1997; open triangles) and Glazebrook \etal
(1995; filled triangles).}
\end{figure}

\begin{figure}
\centerline{
\psfig{figure=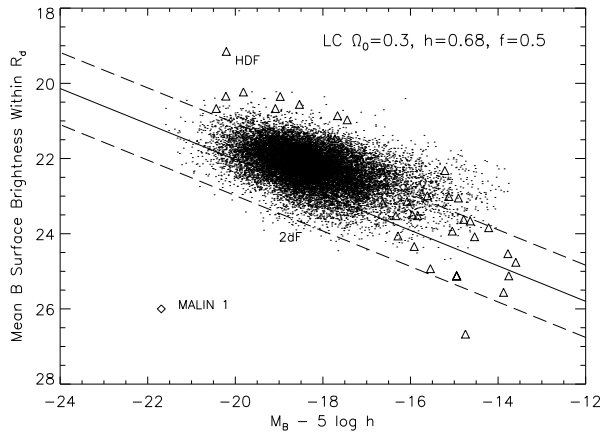,height=6cm,angle=0}}
\caption{\label{fig16} The surface-brightness--magnitude relation. A-model
predictions for the mean $B$-band surface brightness as a function of
magnitude (solid line) at $z=0.4$, which corresponds to the mean redshift of
the HDF data points (open triangles; taken from Driver \etal 1999).  The dots
are the data from Driver \& Cross (2000), derived from the 2dF survey.  The
dashed lines show the 1-$\sigma$ error in the prediction.  We have also
included the location of the anomalous LSB giant Malin 1.}
\end{figure}

\subsection{The Luminosity Function and Surface-Brightness Distribution}

The LF is given by $dn/dM_\lambda = (dn/dM)dM/dM_\lambda$ where $dn/dM$ is the
number density of objects, per unit mass, obtained from equation
(\ref{ps}). An equation for $dM/dM_\lambda$ is derived from the model outputs
assuming the mass-magnitude relationship obeys a power law, which is an
excellent assumption at these scales. For $B$-band magnitudes, we apply the
dust correction formula of Wang \& Heckman (1996) in calculating the LF.

The solid lines in Figure \ref{fig15} are the predictions for the present-day
$B$ and $K$ LFs for the A model.  The horizontal error bars indicate the
1-$\sigma$ error in our mass-magnitude power-law fits.  No color cuts have
been imposed.  The dashed line shows the $B$-band prediction without
correcting for dust.  The dot-dashed line is the $B$-band fit to the 2dF
Survey data (Folkes \etal 1999), and the open squares are data from Zucca
\etal (1997).  The $K$-band data are taken from Gardner \etal (1997; open
triangles) and Glazebrook \etal (1995; filled triangles).

The LF fits are where this model is weakest, as they rely heavily on the
detailed number distribution of galaxies as predicted by the PS formalism,
which is known to be inaccurate.  We find that while the model reasonably
approximates the faint end of the $K$-band data, it slightly underpredicts the
number of galaxies around the knee (near $10^{12}$ $M_\odot$) and overpredicts
the bright end.  We note that the bright end of the observed $K$-band LF is
not presently well constrained, due to sample size and evolutionary effects.
The dust-corrected $B$-band predictions yield reasonable agreement from the
largest galaxies down to about $10^{11}M_\odot$, after which they overshoot
the faint end by a considerable margin, as PS models are known to do.  Given
the shortcomings and limitations of the standard PS formalism, it is quite
remarkable that the model yields as good agreement as it does over a factor of
100 in both number and luminosity.  Somerville \& Primack (1999), using a
modified PS prescription, do achieve slightly better agreement.

Generically, two explanations might resolve the discrepancy at the faint end
in $B$, which is seen to persist even when more sophisticated formalisms than
PS theory are implemented. One possibility is that our models overestimate the
luminosity for a given mass. This would be the case, for example, if supernova
feedback played a significant role in removing the gas. It is unclear to what
extent the energy they release alters the global properties of spirals. If
they serve only to ``puff up'' the gas distribution (Mac Low \& Ferrara 1999),
then our model would remain effectively unchanged. If they expel gas from the
halo, this could alter our predictions.  Inspection of Figure \ref{fig15},
however, reveals that, at low masses, feedback would have to reduce the
predictions by many magnitudes in order to yield agreement. Somerville \&
Primack do find that the inclusion of feedback improves the agreement at the
faint end in $B$, but still returns LF values on the high side. Our TF
predictions, however, match $B$-band data down to $V_c=50$ km s$^{-1}$. Strong
feedback would not only disrupt this agreement for low $V_c$, but also alter
the predictions at other wavelengths. Some recent observations do indicate a
break in the near-IR TF relation at $V_c \sim 90$ km s$^{-1}$ (McGaugh \etal
2000), but these galaxies are found to be extremely gas rich, such that
feedback is not sufficient to expel the gas. Fine tuning by supernovae of the
gas mass available for star formation would be allowed by our model. This
could improve both the TF predictions in $B$ and slightly alleviate the LF
discrepancy, but it appears unlikely that the latter problem can be solved
entirely via feedback.  We point out that the Schmidt law implicitly assumes
supernova feedback which can redistribute gas within the disk and serves to
delay star formation. It is only inflows and outflows which are not accounted
for in our model, and detailed multi-phase ISM simulations seem to indicate
that these are negligible and that spirals may be well-approximated as
``island universes.''  Note also that we may be implicitly assuming some
influence of supernova (or other) feedback in keeping the specific angular
momentum of the baryons equal to that of the halo.

Another possibility is that the model is simply predicting too many small
halos. This may be due in part to the breakdown of the PS formalism in
accurately describing the distribution of halos over these scales, but may
also indicate that the assumed power spectrum itself is supplying too much
power to these scales. Recent evidence suggests that current numerical models
do indeed overpredict the number of dwarf halos surrounding a galaxy such as
the Milky Way by an order of magnitude or more (Moore \etal
2000). Kamionkowski \& Liddle (1999) show that this stark discrepancy can be
addressed through plausible modifications to the power spectrum. This solution
would reconcile the LF predictions with observations, while leaving the TF
predictions untouched.

A very interesting test that is usually overlooked is that of the
surface-brightness--magnitude relation. The solid line in Figure \ref{fig16}
shows our A-model prediction for the mean $B$-band surface brightness within
the disk scale length, $R_d$ [see equation (\ref{Rd})], as a function of
magnitude; the dashed lines are the 1-$\sigma$ bounds. The open triangles are
data from Driver \etal (1999), who used a volume-limited sample derived from
the Hubble Deep Field, and the dots are the data from Driver \& Cross (2000)
derived from the 2dF survey. We have also included the location of the
anomalous LSB giant Malin 1. Although we are plotting model predictions for
$z=0.4$, the mean redshift for the galaxies in the HDF data set, the
brightening of galaxies between $z=0.1$ and 0.4 will have a minor impact on
the outcome of the fit. No color cuts have been imposed. The first thing to
note is that the data correspond to effective surface brightness as defined in
Driver \etal (1999), and therefore include a bulge component which is absent
in our computed central surface brightness for an exponential
disk\footnote{Note that the data points are slightly higher than the Freeman
law (Freeman 1970; Driver \etal 1999).}.  The first striking feature of the
data itself is that there exists a correlation between absolute magnitude and
surface brightness, as first discovered by Binggeli \& Cameron (1991) for the
Virgo population and later confirmed for the field population by Driver \etal
(1999). Gratifyingly enough, our model predicts a similar correlation which
produces a reasonable fit (albeit slightly fainter for the reasons detailed
above and with a scatter only 15\% larger than that {\em currently}
observed). Until more complete surveys can determine the precise scatter of
the surface-brightness--magnitude relation, it is too early to conclude if our
model predicts a scatter that is in fact too large. As already proposed by
Driver \& Cross (2000), this relation offers a physical alternative to the
``botanical'' Hubble-galaxy classification scheme [see Fig. 4 in Driver \&
Cross (2000)], with the x-axis determined by the mass and the y-axis
determined primarily by the angular momentum of the dark halo. Here we confirm
that the x-axis is indeed determined by the mass, but the y-axis is determined
both by the spin of the halo ($\lambda$) {\em and} by the redshift of
formation (i.e., by the initial conditions).  It is also reassuring to find
that our model predicts that LSB giants like Malin-1 should be very rare
objects (about 4-$\sigma$ fluctuations), which agrees well with the observed
paucity of these objects. As was the case for the TF predictions, most other
models investigated (with different cosmogonies and so on) grossly failed to
fit the data. Unlike the TF predictions, however, the surface-brightness
predictions do depend sensitively on $\lambda$ (which fixes the radial scale
length), and thus provide a strong complementary constraint on the models.

\section{Conclusions}

We have constructed a largely analytic disk-galaxy model, built on that of
HJ99, which incorporates such features as the halo formation-redshift
distribution, the joint distribution in spin and peak height, the dependence
on the cosmological model, and star formation with chemical evolution.
Exploring the model's predictions over a range of wavebands, we find that both
parent-halo features and stellar evolution play a role in defining the TF
relation and other disk-galaxy properties; the former fixes the overall
relation and spread, while the latter fine tunes these to observed levels.
Our investigation has yielded a number of specific results:

\begin{itemize}

\item We confirm that the spread in formation redshifts is the primary source
of scatter in the TF relation.  In agreement with previous studies, we predict
a TF relation which generally broadens with decreasing $V_c$ and exhibits
increasingly large scatter towards bluer wavebands (HJ99; van den Bosch 2000;
Somerville \& Primack 1999).

\item We find that $\Omega_0=1$ models are generally too faint to match the
observed TF relation, but contrary to the claim of van den Bosch (2000), this
result does not depend on the inclusion of adiabatic disk contraction,
universal-halo profiles (Navarro, Frenk \& White 1997), or the precise
definition of $V_c$.  We find that successful models tend to favor low values
of $\Omega_0 h \sim$ 0.2, and in particular $\Omega_0 \sim 0.3$ and $h \sim
0.65$ (assuming COBE normalization).\footnote{Recent preliminary results from
the BOOMERANG experiment (Lange \etal 2000) point to larger values of
$\Omega_{b} h^2$. A higher baryon fraction would brighten the TF predictions,
but this could be counteracted by a slightly higher value of $h$ (or by
changing other input parameters so as to still remain within current
constraints), so as to yield a similarly well-fitting model.  Many of the
conclusions presented here are robust with respect to the precise values of
these various parameters.}

\item The inclusion of the spin-parameter distribution, with an
anti-correlation with peak height, tends to act along the TF axis itself and
can therefore slightly reduce the TF scatter in some models, but does not
otherwise have a major impact on the TF relation.  Thus, uncertainties in
obtaining accurate $\lambda$ distributions, such as nonlinear effects, proper
calculation of binding energies, and so on, are likely unimportant.  The spin
distribution is important, however, in determining disk scale lengths and
surface brightnesses, which can provide independent tests of the model.

\item The incorporation of chemical evolution leads to older (younger)
populations which are brighter (fainter), leading to a reduction of TF scatter
of about 0.1 mag in all wavebands.

\item Many models can be found to yield reasonable agreement with the data in
one or two bands, but not in others. Typically, $I$ and $K$ data are the
easiest to fit, since these wavelengths are largely detached from the
star-formation history. Imposing multi-wavelength constraints, however,
provides a strong discriminator among models.

\item Models which yield the best agreement with TF data across {\em all}
wavelengths have cosmological-parameter values in good agreement with current
estimates, and require that most disks from in the range $ 0.5 < z < 2.0$,
with little subsequent accretion, as suggested by Peebles (1999).  In
particular, the reasonable agreement in $B$ seems to rule out very late
collapse.  The $f=0.5$ LC distribution appears to represent a suitable choice
for the range of halo formation redshifts.  Note that this agreement is
obtained with a spread in $z_f$ of $\Delta z_f/(1+ z_f) \sim 0.5$,
considerably larger than that suggested by the results of Eisenstein \& Loeb
(1996) and van den Bosch (2000), demonstrating the impact of regulating
effects from stellar evolution and metal enrichment.

\item Models fitting the TF relation at $z=0$ also appear to obtain the
correct, steeper $B$-band TF relation at $z=1$, and predict a near-IR $z=1$ TF
relation which is essentially identical to the local one at the scales
investigated.

\item Successful models also roughly match the observed surface-brightness
distribution of spirals and reasonably match the observed $B$- and $K$-band
LFs.

\item In general, this success argues that there is probably little room for
stochasticity arising from other complicating effects, such as mergers.
Spiral galaxies appear to behave like island universes (Mac Low \& Ferrara
1999; Peebles 1999).  Small variations to the disk baryon content from such
mechanisms could still be tolerated, though their effects would be largely
confined to the $B$ band. Modeling these weaker effects might serve to improve
the model's prediction for the $B$-band TF relation and LF.

\end{itemize}

There are several features common to many SAMs which have not been
incorporated into our framework, as these do not appear to strongly impact the
properties investigated here.  For example, we adopt an isothermal profile,
rather than the universal form of Navarro, Frenk, \& White (1997). While
observations, as well as some simulations, indicate that real halos do
resemble truncated isothermal spheres, rather than universal profiles
(Sellwood 1999; Dubinski \& Carlberg 1991; Spergel \& Hernquist 1992), our
model is clearly too steep in the core. For our purposes, however, changes to
the profile of a halo with a fixed mass are only relevant inasmuch as they
alter the local gas surface density and thus the SFR. This is similar to the
effect of changing the spin parameter, and we have seen that our model changes
relatively little over the broad range investigated in $\lambda$ and thus
$\Sigma$ (HJ99). Jimenez \etal (1997) developed a similar disk-galaxy model,
but explicitly incorporating the universal profile, and found little
difference from the results of HJ99.  The behavior of an isothermal profile at
the core, however, will have the effect of making younger galaxies too bright
in $B$. As discussed above, this might partially account for the slightly
poorer TF fit and larger predicted scatter in $B$, as well as the discrepancy
with the $B$-band LF.

While the assumption of monolithic disk collapse cannot be strictly accurate,
accounting for adiabatic disk contraction also does not appear essential to
predicting these disk properties. Moreover, adiabatic contraction tends to
make spirals more dark-matter dominated at the core, whereas observations
indicate that they are more baryon dominated (Sellwood 1999).  Our model
depends on an accurate assessment of $V_c$, and not on the details of disk
formation, which do not affect our ultimate results.  Similarly, the inclusion
of a stability threshold criterion for star formation does not appear
essential to our results, and some authors have argued against the importance
of such a threshold (Ferguson \etal 1998).

Another key assumption in this model is that the specific angular momentum of
the baryons is the same as that of the dark matter.  This is a common
assumption of analytic calculations (e.g. Fall \& Efstathiou 1980; Mo, Mao \&
White 1998), but it has been argued from hydrodynamic simulations that this is
not the case; cooling of gas leads to substructure which can couple with the
halo, leading to loss of a sizeable fraction of the baryon angular momentum
(e.g. Navarro \& Benz 1991; Weil, Eke \& Efstathiou; Navarro \& Steinmetz
2000).  The predictions of our model would thus be altered significantly, with
much smaller disks, higher early star formation rates and different colors.
Therefore we need to assume that such angular momentum loss does not occur in
practice.  It is fair to say that the question of how (or indeed if) baryon
angular momentum is conserved is open at present, although there are plenty of
ideas for how the coupling can be prevented.  These include gas ejection from
sub-units by supernovae (Efstathiou 2000), suppression of gas cooling until
the halo is established with a smooth profile (Weil, Eke \& Efstathiou 1998),
prevention of formation of small halos by a cut-off in the power spectrum
(Kamionkowski \& Liddle 1999), or changing the nature of the dark matter
(Spergel \& Steinhardt 1999).  In addition, van Kampen (2000) proposes that
numerical simulations underestimate dynamical friction of the sub-units,
leading to overestimated angular-momentum transfer.

A generic problem is the possibility that our model either over-predicts or
under-predicts the amount of gas which is converted into disk stars, due to
effects such as supernova feedback, the presence of hot x-ray gas, or enhanced
star formation from late infall. Though we have assumed essentially
featureless (in space and time) star formation, the addition or removal of
large amounts of gas via such stochastic mechanisms would increase the scatter
in the TF relation and shatter the agreement obtained here. To the extent that
such processes result in smaller fluctuations to the amount of gas available
for star formation, their effects would be confined primarily to $B$, where
our models are indeed weakest.  Mergers likely play only a very limited
role. It is well known that classic spirals could not maintain their thin-disk
structure in the case of extreme merging.  EL96 found that excluding objects
which had merged with a mass $>20$\% (as opposed to 50\%) of the parent halo,
only reduced the predicted $I$-band scatter by 20\%, suggesting that accretion
does not play a major role, or possibly induces scatter along the
near-infrared TF relation. Either way, this would imply that the TF relation
at higher redshift looks similar to that at $z=0$, and this is precisely what
we find when comparing the $z=0$ and $z=1$ TF relations in $I$ and $K$. Future
work will be aimed at quantitatively assessing the precise impact of these
various other mechanisms, as well as obtaining predictions for the
high-redshift universe.  While our model is a long way from providing a
complete picture of galactosynthesis, our results suggest that future models
addressing spiral properties must invariably incorporate detailed star
formation, as well as initial conditions, and that features such as
exponential disks, little or weak substructure, Schmidt-law star formation,
and little gas inflow or outflow, should be generic predictions of more
sophisticated modelling.

\section*{Acknowledgements} 
We wish to acknowledge the collaboration of A. F. Heavens, whose numerous
contributions have greatly enhanced this work. We also wish to thank L. Wang
for numerous helpful discussions and for providing numerical results pertinent
to our cosmological models.  This work was supported by grants
NSF-AST-0096023, NSF-AST-9900866, NSF-AST-9618537, NASA NAG5-8506, DoE
DE-FG03-92-ER40701 and by a PPARC Advanced Fellowship.

\section*{Appendix}

The luminosity of an arbitrary stellar population can be computed analytically
as follows.  Simple stellar populations (SSPs) are the building blocks of any
arbitrarily complicated population since the latter can be computed as a sum
of SSPs, once the star-formation rate is provided.  In other words, the
luminosity of a stellar population of age $t_0$ (since the beginning of star
formation), in waveband $\lambda$, can be written as:
\begin{equation}
L_{\lambda}(t_0)=\int_{0}^{t_0} \int_{Z_i}^{Z_f} L_{SSP,\lambda}(Z,t_0-t)\,
dZ\, dt
\end{equation}
where the luminosity of the SSP is:
\begin{eqnarray}
&& L_{SSP,\lambda}(Z,t_0-t) = \nonumber \\
&& \int_{M_{min}}^{M_{max}}SFR(Z,M,t)\,
l_{\lambda}(Z,M,t_0-t)\, \frac{dN}{dM}\,dM 
\end{eqnarray}
and $l_{\lambda}(Z,M,t_0-t)$ is the luminosity of a star of mass $M$,
metallicity $Z$, and age $t_0-t$, $Z_i$ and $Z_f$ are the initial and final
metallicities, $dN/dM$ represents the IMF, 
$M_{min}$ and $M_{max}$ are the smallest and largest stellar
mass in the population,
and $SFR(Z,M,t)$ is the star formation rate at the time
$t$ when the SSP is formed. 

The magnitudes for a SSP (normalized to 1 M$_{\odot}$) as a function of age 
and metallicity, for a given photometric band UBVRIJK, are approximated to
within 4\% by:
\begin{equation}
M_{\lambda}=-2.5 \times \sum_{i=0}^{4}\sum_{j=0}^{4} X^i C_{\lambda}(i+1,j+1) Y^j,
\end{equation}
where 
\begin{eqnarray}
X & = & 5.76+3.18  \log{\tau}+1.26  \log^2{\tau}+2.64 
\log^3{\tau} \nonumber \\
& & +1.81  \log^4{\tau} +0.38  \log^5{\tau},
\end{eqnarray}
\begin{eqnarray}
Y & = & 2.0+2.059  \log{\zeta}+1.041  \log^2{\zeta} \nonumber \\ 
& & +0.172 \log^3{\zeta} - 0.042  \log^4{\zeta},
\end{eqnarray}
and
\begin{equation}
\tau=\frac{t}{\rm Gyr}, \;\;\;\;\;\;\;\;\;\;\;\; \zeta = \frac{Z}{Z_\odot}.
\end{equation}

Luminosities are obtained simply from $L_{\lambda}=10^{-0.4*(M_{\odot
\lambda}- M_{\lambda})}$, where $M_{\odot \lambda}=\{5.61, 5.48, 4.83, 4.34,
4.13, 3.72, 3.36, 3.30, 3.28\}$ for $\{U, B, V, R, I, J, H, K, L\}$.  The $i$
and $j$ values appear as exponents of $X$ and $Y$, respectively, and as
indices defining elements of the $C_\lambda$ matrices, given by

\small
\begin{equation}
C_U=\left( \begin{array}{rrrrr}
  -4.738\times 10^{-1}&  4.029\times 10^{-1}& -3.690\times 10^{-1}&  1.175\times 10^{-1}& -1.253\times 10^{-2}\\
  -2.096\times 10^{-1}& -1.743\times 10^{-1}&  1.268\times 10^{-1}& -2.526\times 10^{-2}&  8.922\times 10^{-4}\\
  -1.939\times 10^{-2}&  1.401\times 10^{-2}& -9.628\times 10^{-3}& -1.754\times 10^{-3}&  7.237\times 10^{-4}\\
   2.671\times 10^{-3}& -4.271\times 10^{-4}&  2.470\times 10^{-4}&  2.963\times 10^{-4}& -7.334\times 10^{-5}\\
  -7.468\times 10^{-5}&  6.676\times 10^{-7}&  1.482\times 10^{-6}& -9.594\times 10^{-6}&  2.055\times 10^{-6}
\end{array} \right);
\end{equation}
\begin{equation}
C_B=\left( \begin{array}{rrrrr}
  -8.321\times 10^{-1}&  5.972\times 10^{-1}& -4.818\times 10^{-1}&  1.356\times 10^{-1}& -1.271\times 10^{-2}\\
  -1.223\times 10^{-1}& -2.523\times 10^{-1}&  2.117\times 10^{-1}& -5.309\times 10^{-2}&  3.716\times 10^{-3}\\
  -2.632\times 10^{-2}&  2.468\times 10^{-2}& -2.462\times 10^{-2}&  4.163\times 10^{-3}&  5.054\times 10^{-5}\\
   2.835\times 10^{-3}& -7.906\times 10^{-4}&  9.653\times 10^{-4}& -1.164\times 10^{-5}& -3.800\times 10^{-5}\\
  -7.416\times 10^{-5}&  1.120\times 10^{-6}& -6.707\times 10^{-6}& -5.615\times 10^{-6}&  1.611\times 10^{-6}
\end{array} \right);
\end{equation}
\begin{equation}
C_V=\left( \begin{array}{rrrrr}
  -9.348\times 10^{-1}&  7.376\times 10^{-1}& -5.170\times 10^{-1}&  1.182\times 10^{-1}& -8.023\times 10^{-3}\\
  -9.521\times 10^{-2}& -3.476\times 10^{-1}&  2.379\times 10^{-1}& -4.485\times 10^{-2}&  1.303\times 10^{-3}\\
  -2.437\times 10^{-2}&  4.413\times 10^{-2}& -2.802\times 10^{-2}&  2.202\times 10^{-3}&  5.271\times 10^{-4}\\
   2.625\times 10^{-3}& -2.063\times 10^{-3}&  9.257\times 10^{-4}&  2.378\times 10^{-4}& -8.471\times 10^{-5}\\
  -6.994\times 10^{-5}&  2.842\times 10^{-5}&  7.436\times 10^{-7}& -1.424\times 10^{-5}&  3.050\times 10^{-6}
\end{array} \right);
\end{equation}
\begin{equation}
C_R=\left( \begin{array}{rrrrr}
  -9.755\times 10^{-1}&  8.121\times 10^{-1}& -4.535\times 10^{-1}&  5.553\times 10^{-2}&  3.313\times 10^{-3}\\
  -7.346\times 10^{-2}& -4.000\times 10^{-1}&  1.983\times 10^{-1}& -6.018\times 10^{-3}& -5.621\times 10^{-3}\\
  -2.368\times 10^{-2}&  5.415\times 10^{-2}& -1.968\times 10^{-2}& -5.209\times 10^{-3}&  1.811\times 10^{-3}\\
   2.502\times 10^{-3}& -2.667\times 10^{-3}&  1.376\times 10^{-4}&  8.392\times 10^{-4}& -1.845\times 10^{-4}\\
  -6.690\times 10^{-5}&  4.003\times 10^{-5}&  2.413\times 10^{-5}& -3.043\times 10^{-5}&  5.650\times 10^{-6}
\end{array} \right);
\end{equation}
\begin{equation}
C_I=\left( \begin{array}{rrrrr}
  -1.027\times 10^{0} &  8.951\times 10^{-1}& -3.759\times 10^{-1}& -1.908\times 10^{-2}&  1.690\times 10^{-2}\\
  -4.672\times 10^{-2}& -4.514\times 10^{-1}&  1.411\times 10^{-1}&  4.385\times 10^{-2}& -1.440\times 10^{-2}\\
  -2.379\times 10^{-2}&  6.399\times 10^{-2}& -8.885\times 10^{-3}& -1.425\times 10^{-2}&  3.380\times 10^{-3}\\
   2.414\times 10^{-3}& -3.267\times 10^{-3}& -7.741\times 10^{-4}&  1.532\times 10^{-3}& -3.012\times 10^{-4}\\
  -6.417\times 10^{-5}&  5.166\times 10^{-5}&  4.964\times 10^{-5}& -4.843\times 10^{-5}&  8.607\times 10^{-6}
\end{array} \right);
\end{equation}
\begin{equation}
C_J=\left( \begin{array}{rrrrr}
  -1.106\times 10^{0} &  1.043\times 10^{0} & -1.932\times 10^{-1}& -1.715\times 10^{-1}&  4.364\times 10^{-2}\\
   1.122\times 10^{-2}& -5.240\times 10^{-1}& -6.001\times 10^{-3}&  1.510\times 10^{-1}& -3.228\times 10^{-2}\\
  -2.757\times 10^{-2}&  7.613\times 10^{-2}&  1.935\times 10^{-2}& -3.358\times 10^{-2}&  6.540\times 10^{-3}\\
   2.502\times 10^{-3}& -3.923\times 10^{-3}& -2.956\times 10^{-3}&  2.944\times 10^{-3}& -5.274\times 10^{-4}\\
  -6.439\times 10^{-5}&  6.234\times 10^{-5}&  1.066\times 10^{-4}& -8.371\times 10^{-5}&  1.417\times 10^{-5}
\end{array} \right);
\end{equation}
\begin{equation}
C_K=\left( \begin{array}{rrrrr}
  -1.132\times 10^{0} &  1.296\times 10^{0} & -1.795\times 10^{-1}& -2.391\times 10^{-1}&  5.794\times 10^{-2}\\
   6.838\times 10^{-2}& -6.627\times 10^{-1}& -5.987\times 10^{-2}&  2.115\times 10^{-1}& -4.319\times 10^{-2}\\
  -3.194\times 10^{-2}&  1.009\times 10^{-1}&  3.092\times 10^{-2}& -4.459\times 10^{-2}&  8.454\times 10^{-3}\\
   2.649\times 10^{-3}& -5.499\times 10^{-3}& -3.923\times 10^{-3}&  3.748\times 10^{-3}& -6.625\times 10^{-4}\\
  -6.607\times 10^{-5}&  9.556\times 10^{-5}&  1.332\times 10^{-4}& -1.038\times 10^{-4}&  1.746\times 10^{-5}
\end{array} \right).
\end{equation}

\end{document}